\documentclass[aps,prb,twocolumn]{revtex4}
\usepackage{epsf}

\newcommand{\ybco}{YBa$_2$Cu$_3$O$_{7-\delta}$}

\newcommand{\bscco}{Bi$_2$Sr$_2$CaCu$_2$O$_{8+\delta}$}
\newcommand{\cecuau}{CeCu$_{6-x}$Au$_x$}
\newcommand{\ybrhsi}{YbRh$_2$Si$_2$}

\begin{document}

\title {
Critical properties of the Fermi-Bose Kondo and pseudogap Kondo models:
Renormalized perturbation theory
}

\author{Marijana Kir\'{c}an and Matthias Vojta}
\affiliation{Institut f\"ur Theorie der Kondensierten Materie,
Universit\"at Karlsruhe, Postfach 6980, 76128 Karlsruhe, Germany}
\date{Feb 10, 2004}

\begin{abstract}
Magnetic impurities coupled to both fermionic and
bosonic baths or to a fermionic bath with pseudo\-gap density of
states, described by the Fermi-Bose Kondo and pseudo\-gap Kondo models,
display non-trivial intermediate coupling fixed points
associated with critical local-moment fluctuations and local
non-Fermi liquid behavior.
Based on renormalization group together with a renormalized perturbation expansion
around the free-impurity limit,
we calculate various impurity properties in the vicinity of those intermediate-coupling
fixed points.
In particular, we compute the conduction electron $T$ matrix,
the impurity susceptibility, and the residual impurity entropy,
and relate our findings to certain scenarios of local quantum criticality
in strongly correlated lattice models.
\end{abstract}

\maketitle


\section{Introduction}

Zero-temperature phase transitions in quantum impurity models
have attracted significant interest in recent years in
a number of different contexts.
In correlated electron materials impurities have proven useful for
investigating the complex bulk behavior, by probing the response
to point defects e.g. in a spatially resolved manner by NMR or
scanning-probe microscopy techniques.
In quantum dot systems, impurity models naturally arise in
describing the degrees of freedom on one or multiple dots
together with their coupling to leads.
Finally, the physics of strongly correlated bulk systems can
be mapped to impurity models in the framework of the
dynamical mean-field theory (DMFT) \cite{dmft},
where a local approximation to the lattice electron
self-energy leads to an impurity model supplemented by
a self-consistency condition.

Interesting transitions, where the zero-temperature behavior of
the impurity degrees of freedom changes qualitatively as a
function of some control parameter like coupling constants or gate
voltages, occur in modifications of the familiar Kondo model,
which describes an impurity spin interacting with
conduction electrons of a metallic environment \cite{hewson}.
Whereas the standard Kondo model does not show a phase transition
at a finite value of the coupling,
transitions naturally occur when other effects, which compete
with fermionic Kondo screening, are added.
At the resulting transition point, the Kondo energy scale
is suppressed to zero, and the low-energy physics is dominated by
critical local-moment fluctuations leading to singular thermodynamic properties
and local non-Fermi liquid behavior.
Notably, the physics at these intermediate-coupling fixed points is
different from the well-studied multi-channel Kondo behavior.

In this paper, we will focus on two routes to impurity phase transitions.
The first is represented by the so-called Fermi-Bose Kondo model \cite{bfk,sengupta,bfknew},
describing an impurity spin interacting with both fermionic quasiparticles and
collective spin fluctuations.
The Fermi-Bose Kondo model has recently been analyzed for
the case of a metallic fermion density of states (DOS),
and a gapless bosonic spectrum representing magnetic order parameter fluctuations
at a bulk quantum critical point in $d$ dimensions \cite{bfk,sengupta,bfknew}.
For $d<3$ the model shows an impurity quantum phase transition
between a Kondo-screened phase and a bosonic fluctuating phase with universal
local spin correlations.
Recently, the Fermi-Bose Kondo model has been
proposed \cite{edmft} to describe a ``local quantum phase transition'' in alloys like
\cecuau \cite{schroder}.
This modelling is based on an extended dynamical mean-field
theory \cite{edmft}, where the impurity site is coupled
to both a fermionic and a bosonic self-consistent bath.

A second paradigmatic model is the pseudo\-gap Kondo model \cite{withoff,cassa,bulla,GBI,insi,MVLF}
where the
fermionic bath displays a pseudogap DOS, $\rho_c(\omega)\propto |\omega|^r$.
Such a bath arises in particular in ordered fermionic systems where the order parameter
has nodes at the Fermi surfaces, like unconventional superconductors.
In the pseudogap Kondo model,
screening is suppressed for small Kondo couplings, and a transition
between a Kondo-screened and a free-moment phase results.
The pseudogap Kondo model has been extensively studied by a number of
techniques, in particular using the numerical renormalization group method (NRG).
Recently, it has become clear \cite{insi,MVLF} that $r=0$ and $r=1$ play the
roles of the lower and upper-critical ``dimensions'', respectively,
allowing for different perturbative expansions to access the physics of the
critical point.
On the application side, the pseudogap Kondo effect is relevant for
impurities in $d$-wave high-temperature superconductors, where
indeed non-trivial Kondo-like behavior has been observed connected with the
magnetic moments induced by Zn and Li impurities in \ybco~\cite{bobroff1,bobroff2,julien}.
Furthermore, it has been suggested \cite{kim} that quantum dots coupled to
interacting one-dimensional electron liquids can display pseudogap Kondo
physics.

It is apparent that an analytical description of the properties near the
intermediate-coupling fixed points is highly desirable to obtain a better
understanding of the physics of critically fluctuating moments.
In this paper we shall calculate a number of properties of
the Fermi-Bose Kondo and the pseudogap Kondo models
for parameters in the vicinity of the intermediate-coupling fixed points.
These calculations are based on a perturbative renormalization group
analysis of the models, together with renormalized perturbation
theory -- this allows us to obtain results for observables
in the quantum-critical regime where bare perturbation is known
to diverge.
As in earlier work \cite{bfk,sengupta,science,vbs,bfknew,MVMK},
the expansion is done around the free-moment fixed point
(which can be understood as expansion around the lower-critical
dimension of the phase transitions).
The small parameters controlling the perturbation theory are the
exponents $r$ and $\epsilon=3-d$ occuring in the fermionic and bosonic
bath density of states, respectively.
Critical properties will be computed in a double expansion
in the non-linear couplings of the theory and the scaling dimensions
$r$ and $\epsilon$.
In particular, we will show that the impurity entropy is finite
at the Fermi-Bose critical point, which has consequences for scenarios
of local criticality in the framework of extended DMFT.

The paper is organized as follows.
Sec.~\ref{sec:models} will introduce the impurity models of interest
and describe the properties of the fermionic and bosonic baths.
In Sec.~\ref{sec:RG} we will briefly outline the renormalization group
treatment which is based on a weak-coupling expansion of the
impurity--bath couplings. In particular, all fixed points of the
general pseudogap Fermi-Bose Kondo model will be enumerated and
discussed.
The calculation of observable impurity properties, namely susceptibility,
entropy, and $T$ matrix, 
is presented Secs.~\ref{sec:susc}--\ref{sec:tmatrix}.
When possible, the results are compared with numerical data.
Finally, Sec.~\ref{sec:appl} discusses implications of our study
for scenarios of local criticality in heavy-fermion metals.
Technical details are relegated to the appendices;
in addition we present a large-$N$ calculation of the
impurity entropy for the Bose Kondo model in App.~\ref{app:largen},
which is not restricted small $\epsilon$ and thus
complements the $\epsilon$ expansion of Sec.~\ref{sec:entr}.

Parts of the RG analysis and susceptibility calculations
have been presented in Refs.~\onlinecite{vbs,bfknew,issp,MVMK};
here we generalize these earlier works and derive additional results.
The main new findings concern the impurity entropy, which, to our knowledge,
has not been explicitely evaluated before and is an interesting
property in the DMFT context;
in addition, we present magnetic response results for the pseudogap Kondo model
which can be nicely compared to numerical data.


\section{Models}
\label{sec:models}

It is straightforward to combine the effects of a pseudogap bath and a
competing bosonic interaction onto fermionic Kondo physics
by considering the pseudogap Fermi-Bose Kondo model,
which supersedes the variants discussed above.
The Hamiltonian of the quantum impurity problem can be written as
${\cal H} = {\cal H}_c + {\cal H}_\phi + {\cal H}_{\rm imp}$, with
\begin{eqnarray}
\mathcal{H}_{\text{imp}} &=&
\gamma_0 \hat{S}_{\alpha} \phi_{\alpha} (x\!=\!0) +
 j_0 \hat{S}_{\alpha} (c^\dagger_\nu \sigma^{(\alpha)}_{\nu\mu} c_\mu) (y\!=\!0)
\,.
\label{simp}
\end{eqnarray}
Here, $\hat{S}$ is the operator of the impurity spin, with
$[{\hat S}_\alpha,{\hat S}_\beta]=i\epsilon_{\alpha\beta\gamma}{\hat S}_\gamma$,
and we will restrict our attention to $S=\frac{1}{2}$ in this paper.
The fields $\phi_\alpha(x)$ and $c_\nu(y)$ describe bulk spin fluctuations
and bulk fermionic excitations, respectively, with Hamiltonians ${\cal H}_\phi$ and
${\cal H}_c$ to be discussed in the subsections below.
In particular, the fermionic bath will follow a power law density of states,
\begin{equation}
\rho_c(\omega) \propto |\omega|^r \,,
\end{equation}
which includes the cases of a metal ($r=0$) and a $d$-wave superconductor ($r=1$).
$\nu,\mu$ are spin indices with $\nu=\uparrow,\downarrow$,
$\sigma^{(\alpha)}$ is the vector of Pauli spin matrices,

The impurity part of the Hamiltonian ${\cal H}_{\rm imp}$, contains
couplings to the bosonic and fermionic baths, characterized by
coupling constants $\gamma_0$ and $j_0$.
In the absence of the bosonic bath, $\gamma_0=0$, and for $r=0$
the model (\ref{simp}) reduces to the well-known Kondo model \cite{hewson},
which exhibits screening of the magnetic moment below a characteristic (Kondo)
temperature, $T_K$.
The pseudogap generalization, $r>0$, shows a zero-temperature
phase transition between a free local moment and a screened moment
at a critical coupling $j_0 = j_{0{\rm c}}$.
On the other hand, in the absence of the fermionic bath, $j_0 = 0$,
the Bose-Kondo model describes an impurity moment in an insulating quantum
antiferromagnet~\cite{science,vbs,issp};
in addition it arises in mean-field theories for spin liquids and spin
glasses~\cite{SubirSG}.
Our analysis will be restricted to a model with full rotation symmetry in spin space;
the effect of spin anisotropy for $r=0$ has been discussed in
Ref.~\onlinecite{bfknew}.

\subsection{Fermionic bath}

The fermionic bath consists of non-interacting spin-$\frac{1}{2}$ fermions.
(More precisely, the fermionic self-interaction is assumed to be irrelevant in
the RG sense.)
A convenient continuum representation uses chiral fermions with a linear dispersion
in $(1+r)$ dimensions:
\begin{eqnarray}
{\cal H}_c &=& \int {\rm d}k\,|k|^r (v_F k) c^\dagger_{k\nu} c_{k\nu} \,.
\end{eqnarray}
The field $c_\nu(y=0)$ in Eq.~(\ref{simp}) is simply
the Fourier transform of $c_{k\nu}$,
$c_\nu(y=0) = \int {\rm d}k|k|^r c_{k\nu}$.
The velocity $v_F$ will be set to unity in the following.

We assume particle-hole symmetry of the bath;
in the applicability range of our expansion particle-hole symmetry breaking
is marginal for $r=0$ and irrelevant for $r>0$ and thus does
not change our conclusions.
As will be mentioned in Sec.~\ref{sec:RG},
particle-hole symmetry breaking becomes important
for larger $r$ and/or in the strong-coupling regime.

In $d$-wave superconductors, $r=1$, and the $c$ fermions represent the superconducting
Bogoliubov quasiparticles in the vicinity of the nodes of the $d$-wave gap.

\subsection{Bosonic bath}

The bosonic bath, representing antiferromagnetic
spin-1 collective fluctuations of the host
material, will be described by the O(3)-symmetric
quantum $\phi^4$ theory in $d$ dimensions,
\begin{eqnarray}
\mathcal{H}_\phi = \int{\rm d}^d x \bigg[
\frac{\pi_\alpha^2 + c^2 (\nabla_{x}\phi_{\alpha})^2 +s\phi_\alpha^2}{2} +
\frac{g_0}{4!} \left( \phi_{\alpha}^2 \right)^2 \bigg],
\label{sb}
\end{eqnarray}
where the field $\phi_\alpha(x)$ represents
the local orientation and magnitude of the
(non-conserved) antiferromagnetic order parameter,
and its canonically conjugate momentum is $\pi_\alpha(x)$,
hence $[\phi_\alpha(x),\pi_\beta(x')] = i \delta_{\alpha\beta}\delta^d(x-x')$.
The parameter $s$ tunes a quantum transition
between a paramagnetic phase for $s>s_{\rm c}$ and an
ordered phase breaking the O(3) symmetry at $s<s_{\rm c}$.
$g_0$ is the self-interaction of the $\phi$ quanta,
and $c$ is a velocity, and we will use units such that $c=1$
in the following.
The phase transition of the $\phi^4$ theory can be studied by working
in $(3-\epsilon)$ space dimensions, using a perturbative expansion in
$\epsilon$.

The quantum paramagnet for $s>s_{\rm c}$ is characterized by its spin
gap $\Delta_{\rm s}$ at $T=0$,
and this vanishes as $s$ approaches $s_{\rm c}$ from above as
\begin{equation}
\Delta_{\rm s} \propto (s - s_{\rm c})^{\nu_\phi}
\label{deltacrit}
\end{equation}
where $\nu_\phi$ is the correlation length exponent of the magnetic
bulk transition described by ${\cal H}_\phi$; the dynamical exponent
is $z=1$.
The local two-point $\phi_{\alpha}$ propagator in imaginary time $\tau$
is given by
\begin{eqnarray}
D_\phi(\tau) & = & \langle \phi_{\alpha}(0,\tau) \phi_{\alpha}(0,0)
\rangle_0 \nonumber \\*
&= & T \sum_{\omega_n} \int \frac{{\rm d}^d k}{ (2 \pi)^d}
\frac{e^{-i \omega_n \tau}}{\omega_n^2 + k^2 + m^2}.
\label{eps2}
\end{eqnarray}
Here the momentum $k$ is measured relative to the ordering wavevector
$\bf Q$ of the magnet.
The 0 subscript in the correlator indicates that it is
evaluated to zeroth order in $g_0$. However, we have included
Hartree-Fock renormalizations in the determination of the ``mass'' $m$;
these have been computed in earlier work in the $\epsilon$
expansion \cite{ss}, and we quote some limiting cases to
lowest order in $\epsilon=3-d$:
\begin{equation}
m = \left\{
\begin{array}{ll}
\Delta_{\rm s} & \quad ; \quad s \geq s_{\rm c},~T \ll \Delta_{\rm s} \\
\sqrt{10 \epsilon/33} \, \pi T & \quad ; \quad T \gg
|s-s_{\rm c}|^{\nu_\phi}
\end{array}
\right. .
\label{mass}
\end{equation}
It is useful to define a spectral density, $\rho_\phi(\omega)$, of the Bose field $\phi$.
At $T=0$ it will display a gap of size $\Delta_{\rm s}$;
for $\Delta_{\rm s}=0$ 
it follows
$\rho_\phi(\omega) \propto {\rm sgn}(\omega) |\omega|^{1-\epsilon}$.

For $d<3$ the bosonic self-interaction $g_0$ is relevant in the RG sense
and cannot be neglected.
However, in the RG analysis of the impurity model (\ref{simp}), modifications
from $g_0$ arise only at two-loop order.
Treatments of the Fermi-Bose Kondo model in the context of extended DMFT
\cite{edmft,bfknew} worked with a $g_0=0$ theory -- there the irrelevance of
the bosonic self-interactions was argued to be justified due to the Landau damping
of the spin fluctuations.
In what follows, we will treat both $g_0 = 0$ and $g_0\neq0$ simultaneously,
and display all results for both cases.
In applications, the $g_0\neq0$ case is relevant in the absence of Landau
damping of the spin fluctuations,
i.e., for metallic systems with a small Fermi surface obeying $|{\bf Q}| > 2k_F$
and for pseudogap systems like $d$-wave superconductors \cite{vbs}.


\section{Renormalization group}
\label{sec:RG}

The Fermi-Bose Kondo model (\ref{simp}) can be conveniently analyzed
using standard renormalization group techniques.
We shall assume that the spectrum of the bosonic bath is gapless,
$\Delta_{\rm s}=0$ at $T=0$,
which corresponds to a magnetic bulk quantum critical point.
Then, both fermionic and bosonic baths obey a power law DOS.
The RG can be formulated as expansion in the bath
exponents $\epsilon$ and $r$,
similar to Refs.~\onlinecite{bfk,science,vbs,bfknew}.

It proves convenient to switch from the spin operators
for the impurity in (\ref{simp}) to
an Abrikosov pseudo-fermion representation of the impurity spin $\frac{1}{2}$,
${\vec S} = f^\dagger_\nu {\vec\sigma}_{\nu\mu} f_\mu$.
The required Hilbert space constraint $f_\nu^\dagger f_\nu = \hat{Q} = 1$
will be implemented using a chemical
potential $\lambda\to\infty$,
such that observables $\langle\hat{\cal O}\rangle$ have to be calculated
according to \cite{lambda,costi}
\begin{equation}
\langle\hat{\cal O}\rangle =
\lim_{\lambda\to\infty}
\frac{\langle\hat{Q}\hat{\cal O}\rangle_\lambda}{\langle\hat{Q}\rangle_\lambda} \,,
\label{obs}
\end{equation}
where $\langle\ldots\rangle_\lambda$ denotes the thermal expectation value
calculated using pseudo-fermions in the presence of the chemical potential
$\lambda$.
Clearly, in the limit $\lambda\to\infty$ the term $\langle\hat{Q}\rangle_\lambda$
represents the partition function of the physical sector of the Hilbert
space times $\exp(-\lambda\beta)$.
As detailed in App.~\ref{app:denom}, both numerator {\em and} denominator
of Eq.~(\ref{obs}) have to be expanded in the non-linear couplings to the
required order when calculating observables;
however, the denominator does typically not develop logarithmic singularities
at the marginal dimension.

As an alternative to the pseudo-fermion technique,
the approach of Ref.~\onlinecite{vbs} can be utilized,
which does not employ pseudo-fermions, and is completely equivalent for
the present single-impurity problem.

\subsection{RG transformation}

Some crucial properties of the partition function $Z = {\rm Tr}\,e^{-\beta{\cal H}}$
follow from tree-level power-counting.
The behavior of the fields under a rescaling transformation
determines the scaling dimensions:
\begin{eqnarray}
\text{dim}[\phi_{\alpha}] &=& (d-1)/2 \, ; \quad \text{dim}[c_{\nu}] = (1+r)/2 \,; \nonumber \\
\text{dim}[f_{\nu}] &=& 0.
\label{dimphin}
\end{eqnarray}
The last line refers to the pseudo-fermion representation of the impurity spin.
The above can be used to determine the dimensions of the couplings:
\begin{eqnarray}
\text{dim}[g_{0}] &=& 3-d \,; \nonumber \\
\text{dim}[\gamma_{0}] &=& (3-d)/2 \,; \quad \text{dim}[j_0] = -r \,.
\label{dimgamma}
\end{eqnarray}
Thus, the bosonic bulk interaction $g_0$ is relevant for $d<3$, i.e.,
the bulk critical point will be characterized by a finite fixed point value
of $g$.
The bosonic impurity coupling $\gamma_0$ is a relevant perturbation about the
decoupled impurity fixed point in $d<3$ -- this will yield a {\em stable} fixed
point with finite $\gamma$ \cite{sengupta,science,vbs}.
In contrast, the fermionic coupling $j_0$ is irrelevant, which eventually results in
an {\em unstable} fixed point describing the phase transition in the pseudogap
Kondo model.
The combined effect of $\gamma_0$ and $j_0$ will be captured in the following RG treatment.

\begin{figure}[!t]
\epsfxsize=2.4in
\centerline{\epsffile{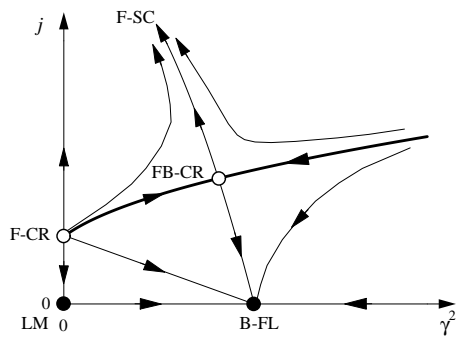}}
\caption{
Schematic RG flow diagram
for the pseudogap Fermi-Bose Kondo model
for small $\epsilon$ and $r$, and $\Delta_{\rm s}\!=\!0$,
deduced from the beta functions (\ref{oneloopbeta}), (\ref{twoloopbeta}).
The horizontal axis denotes the coupling to the bosonic bath, $\gamma$, the
vertical axis denotes the fermionic Kondo coupling, $j$.
The solid dots are (meta)stable fixed points, the open dots
denote critical fixed points, the thick line is the separatrix,
for details see text.
}
\label{fig:flow}
\end{figure}

To one-loop order, the familiar momentum shell method
with ultra-violet (UV) cutoff $\Lambda$ can be employed;
the field-theoretic RG calculation to two-loop order will be sketched
in App.~\ref{app:twoloop}.
Within the momentum shell scheme, dimensionless couplings are introduced as
$g_0 = g \Lambda^\epsilon/S_{d+1}$,
$j_0 = j \Lambda^{-r}$,
$\gamma_0 = \gamma \Lambda^{\epsilon/2}/\widetilde{S}_{d+1}^{1/2}$
with the phase space factors $S_d$, $\widetilde{S}_d$ defined in
Eq.~(\ref{sd}).
Without loss of generality the same value of the UV cutoff can be used for
the fermionic and bosonic subsystems.
The RG for the bulk bosonic theory gives
\begin{equation}
\beta(g) = -\epsilon g + \frac{11 g^2}{6} \,.
\end{equation}
For the impurity part, the one-loop RG result for the beta functions is
\begin{eqnarray}
\beta(\gamma) &=& -\frac{\epsilon\gamma}{2} + \gamma^3 \,, \nonumber\\
\beta(j) &=& rj - j^2 + j\gamma^2  \,,
\label{oneloopbeta}
\end{eqnarray}
the two-loop results can be found in App.~\ref{app:twoloop} in Eq.~(\ref{twoloopbeta}).
Already at one-loop level, a term mixing $j$ and $\gamma$ appears,
which determines the competition between fermionic and bosonic Kondo coupling.
The resulting RG flow in the $j$--$\gamma$ plane
is shown in Fig.~\ref{fig:flow}, the
fixed points will be discussed in the following subsection.

The equations (\ref{oneloopbeta}), (\ref{twoloopbeta}) reduce to the cases of the
pseudogap Kondo model for $\gamma=0$ (Ref.~\onlinecite{withoff}),
the Bose-Kondo model for $j=0$ (Ref.~\onlinecite{vbs}), and the metallic
Fermi-Bose Kondo model with non-interacting bosons for $r=0$
and $g=0$ (Refs.~\onlinecite{sengupta,bfknew}).

\subsection{Fixed points}
\label{sec:FP}

In the following we list the RG fixed points,
obtained from (\ref{twoloopbeta}),
and quote some of their observable properties like the Curie part of
impurity susceptibility $T\chi_{\rm imp}$ and the residual impurity
entropy $S_{\rm imp}$; details of the calculation
appear in Secs.~\ref{sec:susc} and \ref{sec:entr}.

\subsubsection{Local-moment fixed point (LM)}

The decoupled impurity fixed point is simply characterized by
$\gamma^\ast\!=\!j^\ast\!=\!0$.
The susceptibility shows the full Curie moment of a
spin $\frac{1}{2}$, $T\chi_{\rm imp}=1/4$, and the
impurity entropy is $S_{\rm imp} = \ln 2$.

\subsubsection{Fermionic strong-coupling fixed point (F-SC)}

For large fermionic Kondo coupling $j$,
the bosonic part $\gamma$ is irrelevant, and the model reduces to the fermionic
pseudogap Kondo model which shows strong-coupling phases with screening
($j^\ast=\infty$, $\gamma^\ast=0$).
This regime is not accessible within the present perturbation expansion,
and we just quote the known results for the Kondo model.
For the metallic case, $r=0$, there is a line of strong-coupling fixed points
parametrized by particle-hole asymmetry, along this line
$T\chi_{\rm imp}=0$ and $S_{\rm imp} = 0$.
In contrast, for $r>0$ two different strong-coupling fixed points exist \cite{GBI}.
The particle-hole symmetric one is unstable w.r.t. particle-hole symmetry breaking
and only accessible for $r<1/2$, it does
not display complete screening: the susceptibility follows
$T\chi_{\rm imp}=r/8$, and the impurity entropy is $S_{\rm imp} = 2 r \ln 2$.
In contrast, the asymmetric strong-coupling fixed point is stable and
exists for all $r>0$, with
$T\chi_{\rm imp}=0$ and $S_{\rm imp} = 0$.

\subsubsection{Bosonic fluctuating fixed point (B-FL)}

In the absence of a fermionic bath, the Bose Kondo problem
shows interesting behavior:
it flows to a stable intermediate-coupling fixed point,
characterized by
\begin{eqnarray}
{\gamma^\ast}^2 &=& \left\{
\begin{array}{ll}
\frac{\epsilon}{2} + \epsilon^2 \big( \frac{29}{121} - \frac{5\pi^2}{132}\big) + {\cal O}(\epsilon^3)
  & \mbox{for } g_0\neq0 \\[8pt]
\frac{\epsilon}{2} + \frac{\epsilon^2}{4} + {\cal O}(\epsilon^3)
  & \mbox{for } g_0=0
\end{array}
\right. \nonumber \\[8pt]
j^\ast&=&0 \,.
\end{eqnarray}
At this fixed point the fermionic Kondo coupling $j$ is irrelevant.
The intermediate-coupling nature leads to universal local-moment
fluctuations \cite{science,vbs}, with
$T\chi_{\rm imp} > 0$ and $0<S_{\rm imp}<\ln 2$.
The anomalous dimension of the auxiliary impurity fermionic field
(see also App.~\ref{app:twoloop}) is
\begin{equation}
\eta_f = \left\{
\begin{array}{ll}
\frac{3\epsilon}{8} -\frac{3\epsilon^2}{8}
   \left( \frac{5}{242} + \frac{5\pi^2}{66} \right) + {\cal O}(\epsilon^3)
  & \mbox{for } g_0\neq0 \\[8pt]
\frac{3\epsilon}{8} + {\cal O}(\epsilon^3)
  & \mbox{for } g_0=0
\end{array}
\right. \,.
\end{equation}
$\eta_f$ describes the decay of the two-point correlator of the $f$ fermions.
In the Kondo model (\ref{simp}) this propagator is not directly measurable;
however, it has been argued \cite{soc} that mobile fermionic quasiparticles in a
critical antiferromagnet are described by a very similar theory,
provided that their momentum is close to a minimum or van-Hove point of the
band structure, such that the dispersion is quadratic and can be neglected.
Then, $\eta_f$ is the exponent of the electron Green's function as
measured in photo\-emission:
\begin{equation}
G_e \propto \frac{1}{(\omega-\epsilon_0)^{1-\eta_f}} \,,
\end{equation}
where $\epsilon_0$ is the (renormalized) energy of the quasiparticle.
This situation, which can be regarded as the bosonic analogue of the
familiar x-ray edge problem \cite{ssbose}, applies to electrons in Kondo insulators
at a magnetic critical point, and could also be relevant for holes near the
anti-nodal points in high-temperature superconductors \cite{soc}.

As noted earlier \cite{vbs,bfknew} the B-FL fixed point is unstable
w.r.t. breaking of the spin rotation symmetry.
Further, for $\epsilon\to 0$ it merges with LM.

\subsubsection{Fermionic critical fixed point (F-CR)}

The pure fermionic problem with a pseudogap, $r>0$, features a critical fixed
point with
\begin{equation}
j^\ast = r + \frac{r^2}{2} + {\cal O}(r^3)\,, \quad \gamma^\ast=0
\end{equation}
which controls the transition between the decoupled and the Kondo-screened
phases.
As detailed below, we have $0 < T \chi_{\rm imp} < 1/4$ and $S_{\rm imp}>\ln 2$.
The anomalous field dimension is given by
\begin{equation}
\eta_f = \frac{3r^2}{8} + {\cal O}(r^3) \,.
\end{equation}
As the fixed point is infrared unstable, we can discuss the
scaling dimension of the leading relevant operator, which
is just the inverse correlation length exponent of
the LM -- F-SC transition.
Expanding the beta function around the fixed point value yields:
\begin{equation}
\frac{1}{\nu} = r - \frac{r^2}{2} +  {\cal O}(r^3)\,.
\label{nuz-fcr}
\end{equation}
This exponent describes the vanishing of the characteristic energy
scale $T^\ast$ when the system is tuned through the transition,
i.e., $T^\ast \propto |j_0 - j_{0{\rm c}}|^{\nu}$.

The F-CR fixed point is unstable w.r.t a finite coupling to gapless bosons,
therefore it cannot be reached for $\gamma_0\neq 0$ if $\Delta_{\rm s}=0$.
The flow diagram, Fig.~\ref{fig:flow} shows that F-CR can be understood
as a multicritical fixed point;
consequently, universal one-parameter scaling in its vicinity
is only realized for $\gamma_0=0$.
For $r\to 0$, the F-CR fixed point merges with LM.

NRG calculations \cite{GBI}
have established that the fixed-point structure of the pseudogap Kondo
problem changes for larger $r$, and two different critical fixed points
occur for $r>r^\ast\approx 0.375$.
The particle-hole symmetric critical fixed point, present at small $r$,
disappears for $r\geq \frac{1}{2}$.
Notably, the particle-hole symmetric and asymmetric critical fixed points share
many common properties \cite{GBI,insi,MVLF}.
In any case, the physics at large $r$ is inaccessible within the present
weak-coupling expansion;
we will restrict our attention to the range of $0<r<\frac{1}{2}$ in the following.

\subsubsection{Fermi-Bose critical fixed point (FB-CR)}

Finally, the FB-CR fixed point controls the transition between F-SC and B-FL,
i.e., between the phase with fermionic screening and the bosonic fluctuating phase.
For interacting bosons, $g_0\neq 0$, we have
\begin{eqnarray}
{\gamma^\ast}^2 &=&
  \frac{\epsilon}{2} + \epsilon^2 \left( \frac{111}{968}-\frac{5\pi^2}{132}\right)
  - \frac{\epsilon r}{2} - \frac{r^2}{2} + {\cal O}(\epsilon,r)^3\,, \nonumber\\
j^\ast&=& r + \frac{\epsilon}{2} - \epsilon^2 \left(\frac{5}{484}+\frac{5\pi^2}{132}\right)
  + {\cal O}(\epsilon,r)^3
\end{eqnarray}
whereas for $g=0$ we find
\begin{eqnarray}
{\gamma^\ast}^2 &=& \frac{\epsilon}{2} + \frac{\epsilon^2}{8} - \frac{\epsilon r}{2} -
\frac{r^2}{2} + {\cal O}(\epsilon,r)^3 \,, \nonumber\\
j^\ast          &=& r + \frac{\epsilon}{2} +{\cal O}(\epsilon,r)^3 \,.
\end{eqnarray}
The fixed point is again characterized by non-trivial universal
values of $T\chi_{\rm imp}$ and $S_{\rm imp}$ depending on $r$ and $\epsilon$
only, in particular for $r=0$ we have $0<S_{\rm imp}<\ln 2$.
The anomalous field dimension is
\begin{equation}
\eta_f = \left\{
\begin{array}{ll}
\frac{3\epsilon}{8} - \frac{3\epsilon^2}{8}
   \left( \frac{5}{242} + \frac{5\pi^2}{66} \right) + {\cal O}(\epsilon,r)^3
  & \mbox{for } g_0\neq0 \\[8pt]
\frac{3\epsilon}{8} + {\cal O}(\epsilon,r)^3
  & \mbox{for } g_0=0
\end{array}
\right. \,.
\end{equation}
The inverse correlation length exponent of the B-FL -- F-SC transition is
\begin{equation}
\frac{1}{\nu} = \left\{
\begin{array}{ll}
r+\frac{\epsilon}{2} - \epsilon^2 \big( \frac{34}{363} + \frac{5\pi^2}{132}\big) - \frac{4\epsilon r}{9} - \frac{19r^2}{27}
  & \mbox{for } g_0\neq0 \\[8pt]
r+\frac{\epsilon}{2} - \frac{\epsilon^2}{12} - \frac{4\epsilon r}{9} - \frac{19r^2}{27}
  & \mbox{for } g_0=0
\end{array}
\right.
\label{nuz-fbcr}
\end{equation}
up to terms of order ${\cal O}(\epsilon,r)^3$.
Note that this exponent describes the behavior if the impurity
model is tuned off criticality by varying $j_0$ or $\gamma_0$,
while maintaining criticality of the bulk magnet (!), $\Delta_{\rm s}=0$.
If, in contrast, the bosonic bath is tuned away from its critical
point, then the physics is dominated by $\Delta_{\rm s}$ (\ref{deltacrit}),
as discussed in Ref.~\onlinecite{vbs} (see also Sec.~\ref{sec:bvsb} below).
The FB-CR fixed point only exists for $\epsilon>0$.

The results of Refs.~\onlinecite{GBI,MVLF} suggest that the fixed point
character of FB-CR changes for larger $r$ similar to that of
the F-CR fixed point.
In contrast, the properties of the bosonic part are likely
smooth as function of $\epsilon$ for $0<\epsilon<2$ ($1<d<3$),
as the $(1+\epsilon)$ expansion of Ref.~\onlinecite{impnlsm} yields
universal properties similar to the $(3-\epsilon)$ expansion \cite{vbs};
this is further supported by the large-$N$ theory
presented in Ref.~\onlinecite{vbs} and App.~\ref{app:largen}.

\subsection{Phase diagram}

The schematic RG flow from this small $(\epsilon, r)$ analysis is shown in
Fig.~\ref{fig:flow}.
The main conclusion is that the fermionic and bosonic Kondo couplings
compete, i.e., the coupling of an impurity to collective spin fluctuations
suppressed fermionic Kondo screening.
This can be an important aspect for impurity moments in strongly correlated
materials \cite{MVMK}.

We expect the general structure of the phase diagram
to be valid for all values $0<\epsilon<2$ and $r>0$.
This is suggested by numerics for the fermionic part \cite{bulla,GBI} and
various large-$N$ methods \cite{vbs,MVMK};
for $r>\frac{1}{2}$ particle-hole asymmetry is needed to reach the
F-SC fixed point as noted above.
For the case of a metallic fermionic bath, $r=0$, the F-CR fixed point is
absent, and the phase transition line originates in the $\gamma=j=0$
point, see Refs.~\onlinecite{sengupta,bfknew}.

\subsection{Tuning bulk vs. boundary criticality}
\label{sec:bvsb}

So far the analysis was restricted to $s=s_{\rm c}$, i.e., the bulk quantum
critical point of the host magnet;
here we have seen that tuning the couplings $j_0$ and $\gamma_0$ can
lead to boundary quantum phase transitions.

It is instructive to also discuss the physics away from the bulk critical
point, and in particular the behavior of the impurity properties
upon tuning the {\em bulk} magnet through its ordering transition.
The behavior will of course depend on whether the values of
the impurity couplings, $j_0$ and $\gamma_0$, place the impurity
problem at $s=s_{\rm c}$ into the F-SC or B-FL phases, or on the phase
transition line, Fig.~\ref{fig:flow}.

\subsubsection{$r=0$}

For the metallic case, $r=0$,
a finite host spin gap, corresponding to $s>s_{\rm c}$, always
leads to an impurity screened by fermionic quasiparticles.

Then, for large $j_0$ there is no qualitative change in the impurity properties
when tuning $(s-s_{\rm c})$ through zero, because
for $s<s_{\rm c}$ the physics is described by a screened Kondo impurity
in a weak magnetic field \cite{hewson}.
In contrast, for small $j_0$ the low-energy behavior of the impurity
changes from F-SC for $s>s_{\rm c}$ to B-FL at $s=s_{\rm c}$,
and for $s<s_{\rm c}$ the moment is polarized by the exchange field,
with an effective magnetization $\propto (s_{\rm c}-s)^{\nu_\phi \eta_\chi/2}$
(Ref.~\onlinecite{vbs}).
Only for $j_0$, $\gamma_0$ placing the impurity problem on the
transition line in Fig.~\ref{fig:flow}, the impurity
is driven critical by driving the bulk to the critical point.

\subsubsection{$r>0$}

In the pseudogap situation, $r>0$,
the presence of a finite host spin gap, $\Delta_{\rm s}>0$, still allows
a transition between LM and F-SC.
This can be understood in a two-step RG procedure:
For UV cutoff energies larger than $\Delta_{\rm s}$ the above RG can be applied,
generating a flow of $j$.
Once the UV cutoff reaches $\Delta_{\rm s}$, the RG flow of $\gamma$ is terminated.
The remaining RG procedure then effectively treats a model with fermionic bath
only, but the flow of $j$ starts at $j(\Delta_{\rm s})$ generated in the first
part of the RG.
Thus for $r>0$ there exists a transition between LM and F-SC as function of
$j_0$, controlled by the F-CR fixed point.

We can now discuss the behavior when tuning the bulk magnet through its critical point.
For large $j_0$ the impurity is screened, and its properties remain essentially
unchanged upon tuning $(s-s_{\rm c})$ through zero as above.
For very small $j_0$ the low-energy behavior of the impurity
changes now from LM (!) for $s>s_{\rm c}$ to B-FL at $s=s_{\rm c}$, and at $s<s_{\rm c}$
the moment is polarized.
Interestingly, there is now a significant intermediate range of $j_0$
values where the impurity is screened for large $\Delta_{\rm s}$,
then becomes uncreened for smaller $\Delta_{\rm s}$ (Ref.~\onlinecite{MVMK})
-- this transition is the one controlled by F-CR --
and at $s\leq s_{\rm c}$ the behavior again is determined
by B-FL \cite{vbs}.

Summarizing this discussion, for the single-impurity model
it is clear that, for generic
values of the impurity couplings $j_0$ and $\gamma_0$, tuning
the bulk magnet critical does {\em not} drive the impurity
problem critical;
coincidence of the two critical points requires additional fine
tuning of parameters.


\section{Magnetic susceptibility}
\label{sec:susc}

The linear response functions to an applied field have
been discussed for the purely bosonic model in Ref.~\onlinecite{vbs};
here we supplement this by calculating the fermionic contributions
and collecting the results for the different fixed points.
Also, we briefly discuss hyperscaling properties with
respect to a local field near the critical fixed points.

The external field is coupled to the bulk magnetic degrees of freedom
using the following replacement in ${\cal H}_\phi$,
\begin{eqnarray}
\pi_{\alpha}^2 &\rightarrow& (\pi_{\alpha} -i
\epsilon_{\alpha\beta\gamma} H_{\text{u}\beta} (x) \phi_{\gamma} )^2 \,,
\label{par1}
\end{eqnarray}
while we simply add field terms to ${\cal H}_c$ and $\mathcal{H}_{\text{imp}}$,
\begin{eqnarray}
&&- H_{\text{u}\alpha}(y) (c^\dagger_\nu \sigma_{\nu\mu}^\alpha c_\mu)(y) ~\mbox{and} \nonumber \\
&&- H_{\text{imp},\alpha} {\hat S}_{\alpha} \,.
\label{par2}
\end{eqnarray}
The bulk field $H_{\text{u}}$ varies
slowly as function of the space coordinate, and $H_{\text{imp}}$ is the magnetic
field at the location of the impurity.

With these definitions, a spatially uniform field applied
to the whole system corresponds to $H_{\rm u} = H_{\rm imp} = H$.
Response functions can be defined from second derivatives of the thermodynamic
potential, $\Omega = - T \ln Z$, in the standard way \cite{vbs}:
$\chi_{\rm{u},\rm{u}}$ measures the bulk response to a field applied
to the bulk, $\chi_{\rm{imp},\rm{imp}}$ is the impurity response to
a field applied to the impurity, and $\chi_{\rm{u},\rm{imp}}$
is the cross-response of the bulk to an impurity field.
Then the impurity contribution to the total susceptibility is defined as
\begin{eqnarray}
\chi_{\rm imp}(T)
= \chi_{\rm imp,imp} + 2 \chi_{\rm u,imp} + (\chi_{\rm u,u} - \chi_{\rm u,u}^{\rm bulk})
\,,
\end{eqnarray}
where $\chi_{\rm u,u}^{\rm bulk}$ is the susceptibility of the bulk system in
absence of impurities.
The local impurity susceptibility is given by
\begin{equation}
\chi_{\rm loc}(T) = \chi_{\rm imp,imp} \,,
\label{chiloc}
\end{equation}
which is equivalent to the zero-frequency impurity
spin autocorrelation function.

\subsection{Local susceptibility}

The local susceptibility $\chi_{\rm loc}(T)$ at an intermediate-coupling
fixed point is expected to follow a power law,
\begin{equation}
\lim_{T\to 0} \chi_{\rm loc}(T) \propto \frac{1}{T^{1-\eta_\chi}} \,.
\end{equation}
This relation defines an anomalous exponent, $\eta_\chi$, which
controls the anomalous decay of the two-point correlations
of the impurity spin.
(In Refs.~\onlinecite{vbs,issp} this exponent was labelled $\eta'$.)

Following the standard scheme \cite{bgz},
$\eta_\chi$ is calculated by introducing a $\chi_{\rm loc}$ renormalization
factor, $Z_\chi$, evaluating the lowest-order diagrams for $\chi_{\rm loc}$,
and demanding the resulting renormalized expression to be free of poles.
From the two-loop expression for $Z_\chi$, given in App.~\ref{app:twoloop},
we find $\eta_\chi = 2 (\gamma^2-\gamma^4) + j^2$.
Evaluating this at the RG fixed points yields
\begin{eqnarray}
 \mbox{F-CR}:~\eta_\chi &=& r^2 + {\cal O}(r^3) \,, \\
\mbox{FB-CR}:~\eta_\chi &=& \epsilon - \epsilon^2
   \left( \frac{5}{242} + \frac{5\pi^2}{66} \right) + {\cal O}(\epsilon,r)^3 \,,  \nonumber\\
 \mbox{B-FL}:~\eta_\chi &=& \epsilon - \epsilon^2
   \left( \frac{5}{242} + \frac{5\pi^2}{66} \right) + {\cal O}(\epsilon^3)\,,  \nonumber
\end{eqnarray}
where the last two lines are for interacting bosons.
We see no reason that the $\eta_\chi$ values for the FB-CR and
B-FL fixed points should be equal to all orders.
In contrast, for non-interacting bosons, $g=0$, it has been shown
in Refs.~\onlinecite{vbs,bfknew} that all higher-order corrections to
$\eta_\chi$ vanish identically at a fixed point with finite $\gamma^\ast$,
thus
\begin{eqnarray}
\label{exact_etachi}
\mbox{FB-CR}:~\eta_\chi &=& \epsilon \,, \\
 \mbox{B-FL}:~\eta_\chi &=& \epsilon  \nonumber
\end{eqnarray}
to all orders in perturbation theory.
To the order we have worked here we find $\eta_f = 3 \eta_\chi / 8$,
but this does not hold to higher orders.

\subsection{Hyperscaling}

The structure of the RG implies that the impurity correlations
obey certain hyperscaling properties at the non-trivial fixed points,
including $\omega/T$ scaling in dynamical quantities \cite{book}.
For instance, the dynamic local susceptibility at criticality obeys
\begin{equation}
\chi''_{\rm loc}(\omega,T)
= \frac {{\cal B}_1} {\omega^{1-\eta_{\chi}}} \, \Phi_1 \!\left(\frac{\omega}{T}\right)
\,,
\label{imchi}
\end{equation}
where $\Phi_1$ is a universal crossover function (for the particular fixed point),
and ${\cal B}_1$ is a non-universal prefactor.
The asymptotic behavior of $\Phi_1(x)$ is
\begin{equation}
\Phi_1(x) = \left\{
\begin{array}{ll}
x^{1-\eta_\chi} & \mbox{for } x\to 0 \\[8pt]
{\rm const}     & \mbox{for } x\to\infty
\end{array}
\right. \,.
\end{equation}
A similar scaling form applies for the $T$ matrix, see below.

\begin{figure}
\epsfxsize=3.4in
\centerline{\epsffile{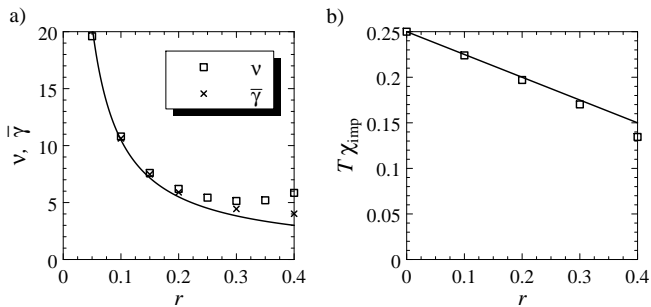}}
\caption{
Comparison of numerical data for the critical point
of the pseudogap Kondo model (F-CR), obtained with NRG,
to the analytical results of the renormalized perturbation theory.
a) Correlation length exponent $\nu$ and susceptibility exponent
$\bar{\gamma}$ (\ref{bargammadef}).
The analytical results are in (\ref{nuz-fcr}) and (\ref{gammares}):
to two-loop order, both exponents are equal, and shown by the solid line.
The symbols are NRG data from Refs.~\protect\onlinecite{bulla,insi}.
b) Curie prefactor $C_{\rm imp}$ of the impurity susceptibility (\ref{fract}).
Solid: analytical result (\ref{tchires});
squares: NRG data taken from Ref.~\protect\onlinecite{GBI}.
}
\label{fig:cmp}
\end{figure}

In addition, for the critical fixed points F-CR and FB-CR we
can also consider scaling properties away from, but close to, criticality.
For details in the context of the pseudogap Kondo model we refer the reader
to Ref.~\onlinecite{insi}.
In particular, the static local susceptibility follows
\begin{equation}
\chi_{\rm loc}(T)
= \frac {{\cal B}_2} {T^{1-\eta_{\chi}}} \, \Phi_2 \!\left(\frac{T^{1/(\nu)}}{j_0-j_{0{\rm c}}}\right)
\,.
\label{chiscal2}
\end{equation}
Using hyperscaling, the results for the correlation length exponent $\nu$, Sec.~\ref{sec:RG},
and for the anomalous exponent $\eta_\chi$ from above
are sufficient to determine all critical exponents associated with
a local magnetic field \cite{insi}.
In particular, the $T\to 0$ local susceptibility away from criticality
obeys:
\begin{eqnarray}
\chi_{\rm loc}(j_0\!>\!j_{0{\rm c}})  &\propto& (j_{0}-j_{0{\rm c}})^{-\bar{\gamma}} \,,~
\bar{\gamma} = \nu \,(1-\eta_\chi) \,,\nonumber\\
T\chi_{\rm loc}(j_0\!<\!j_{0{\rm c}}) &\propto& (j_{0{\rm c}}-j_{0})^{\bar{\gamma'}} \,,~
\bar{\gamma'} = \nu \, \eta_\chi \,.
\label{bargammadef}
\end{eqnarray}
With these definitions, the exponents $\bar{\gamma}$ and $\bar{\gamma'}/2$
coincide with the exponents $\gamma$ and $\beta$ defined
in Ref.~\onlinecite{insi}.
The perturbative expressions for the exponents in Eq.~(\ref{bargammadef}) are
\begin{eqnarray}
 \mbox{F-CR}:~\bar{\gamma} &=& \frac{1}{r} + \frac{1}{2} + {\cal O}(r) \,, \label{gammares} \\
              \bar{\gamma'}&=& r + {\cal O}(r^2) \,, \nonumber \\
\mbox{FB-CR}:~\bar{\gamma} &=& \frac{2}{2r+\epsilon} +{\cal O}(1) \,, \nonumber\\
              \bar{\gamma'}&=& \frac{2\epsilon}{2r+\epsilon} + {\cal O}(\epsilon,r) \nonumber \,.
\end{eqnarray}
At this order the results for $g=0$ and $g\neq 0$ are identical;
the next-to-leading order contributions at the FB-CR fixed point
can also be found from (\ref{nuz-fbcr}).
The results for the F-CR fixed point can be compared with numerical results \cite{insi},
with excellent agreement, e.g.,
$\bar{\gamma}(r=0.1) = 10.63 \pm 0.02$ from NRG,
see also Fig.~\ref{fig:cmp}.

As proposed earlier \cite{GBI,insi}, $\lim_{T\to 0}T\chi_{\rm loc}$ can serve as an order
parameter, as it vanishes continuously as $j_0\to j_{0{\rm c}}$ from below,
and is zero for $j_0>j_{0{\rm c}}$, i.e., in the F-SC phase.

\subsection{Impurity susceptibility}

The quantity $\chi_{\rm imp}$ measures the impurity
contribution to the total uniform susceptibility.
Importantly, it does not
acquire an anomalous dimension at an intermediate-coupling fixed point\cite{ss},
because it is a response function associated to the conserved
quantity $S_{\rm tot}$.
Thus we expect a Curie law
\begin{equation}
\lim_{T\to 0} \chi_{\rm imp}(T) = \frac{C_{\rm imp}}{T} \,,
\label{fract}
\end{equation}
where the prefactor is in general a non-trivial universal constant different
from the free-impurity value $S(S+1)/3$.
Apparently, Eq.~(\ref{fract}) can be interpreted as the Curie response of a
fractional effective spin \cite{science}.

Since the present expansion is done around the free-impurity fixed point of the
$S=\frac{1}{2}$ impurity, the result will always have the form
$T \chi_{\rm imp} = \frac{1}{4} + \Delta (T \chi_{\rm imp})$.
Technically, the absence of an anomalous dimension implies that
no poles in $\epsilon$ or $r$ will appear in the perturbative
expression for $\chi_{\rm imp}$, i.e., all poles from the different
contributions have to cancel.
Importantly, such pole-free contributions can also arise from
the denominator in Eq.~(\ref{obs}), see App.~\ref{app:denom}.

At this point a reference to the two-channel Kondo problem is in order,
where it is known that $\chi_{\rm imp}$ {\em does} display an anomalous
power law.
This does not contradict the above statement, however, in the two-channel model
the prefactor $C_{\rm imp}$ of Eq.~(\ref{fract}) actually vanishes
due to a ``compensation'' effect \cite{barzykin},
and the leading low-temperature behavior arises from corrections
to scaling \cite{AL,gtheorem}.

Proceeding with our calculation, we first focus on the fermionic part.
The analysis of $\chi_{\rm loc}$ has shown that the first
corrections to free-moment behavior arise at order $j_0^2$.
This is different for $\chi_{\rm imp}$:
there is a single diagram proportional to $j_0$
which contributes to $\chi_{\rm u,imp}$:
\begin{equation}
\chi_{\rm u, imp} = - j_0 \frac{1}{4 T^2}\int {\rm d}k |k|^r \frac{\cosh^{-2}(k/2T)}{4}
\end{equation}
Evaluating the integral in the limit of infinite UV cutoff and taking $r\to 0$ gives:
\begin{equation}
\Delta (T\chi_{\rm imp})^{({\rm f})} = - \frac{j}{4}
\end{equation}
where we have already expressed the result in terms of the renormalized coupling $j$,
$j_0 = j\mu^{-r}$,
and used $(T/\mu)^r = 1 + {\cal O}(r)$.

The bosonic contributions have been discussed in Ref.~\onlinecite{vbs};
here we only quote the result for completeness.
Importantly, the impurity susceptibility is not a well-defined quantity
for non-interacting bosons ($g=0$) because the application of a uniform field
would lead to negative bosonic mode energies -- this yields an infrared
singular magnetic response of the bulk system, and a similar infrared
singularity of $T\chi_{\rm imp}$.
For interacting bosons, the lowest-order term gives \cite{vbs}
\begin{equation}
\Delta (T\chi_{\rm imp})^{({\rm b_{int}})} =
\frac{1}{4} \sqrt{\frac{33}{10\epsilon}} \, \gamma^2
\end{equation}
where the Hartree-Fock renormalization of the mass (\ref{mass})
is required to remove the infrared divergence of the integral.

With these perturbative results, we can finally evaluate the
impurity susceptibility at the intermediate-coupling fixed points,
to lowest non-trivial order.
For the bosonic part, the order $\epsilon$ contribution can
also be identified \cite{vbs}.
\begin{eqnarray}
 \mbox{F-CR}:~T\chi_{\rm imp} &=& \frac{1}{4} (1 - r) + {\cal O}(r^2) \,, \label{tchires} \\
\mbox{FB-CR}:~T\chi_{\rm imp} &=& \frac{1}{4}
   \left( 1 + \sqrt{\frac{33 \epsilon}{40}} - \frac{9\epsilon}{4} - r\right) \nonumber\\
  &~&~~~~~~~~~~~~~~~~~~~ + {\cal O}(\epsilon^{3/2},\epsilon^{1/2}r) \,,  \nonumber\\
 \mbox{B-FL}:~T\chi_{\rm imp} &=& \frac{1}{4}
   \left( 1 + \sqrt{\frac{33 \epsilon}{40} } - \frac{7\epsilon}{4} \right)
   + {\cal O}(\epsilon^{3/2}) \nonumber \,.
\end{eqnarray}
Notably, our result for the F-CR fixed point is
in excellent agreement with the NRG of Ref.~\onlinecite{GBI},
see Fig.~\ref{fig:cmp}.

For the bosonic case no numerical results for small $\epsilon$ are available;
calculations \cite{troyer} for $\epsilon=1$ ($d=2$) have found a value for $T\chi_{\rm imp}$
which is very close to the free-moment value $\frac{1}{4}$.
We believe that this coincidence is due to large corrections from higher
orders of the $\epsilon$ expansion;
the fact that our result for the fermionic F-CR fixed point agrees well with numerics
strongly supports that the renormalized perturbation expansion employed here is
the correct method to evaluate the impurity susceptibility
(which had been questioned in Ref.~\onlinecite{sushkov}).


\section{Impurity entropy}
\label{sec:entr}

In general, zero-temperature critical points in quantum impurity models
can show a finite residual entropy -- this is in contrast
to bulk quantum critical points where the entropy usually vanishes
at $T=0$, but features an unconventional power law,
$S(T) \propto T^x$.

For the models at hand, the impurity contribution to the
low-temperature entropy is obtained by a perturbative evaluation
of the thermodynamic potential and taking the temperature
derivative.
In the present expansion the result has the form
$S_{\rm imp}(T=0) = \ln (2S+1) + \Delta S_{\rm imp}$ with $S=\frac{1}{2}$.

It is worth emphasizing that the aim of the $\epsilon$ expansion
for the entropy is not to capture {\em singular} corrections
(as is the case for observables which develop anomalous power laws),
but to determine the power series for the coefficient of an observable with
{\em regular} behavior.
As shown below, fully universal quantities can be obtained by
correctly interpreting the result of the perturbative calculation.
The situation is similar to the evaluation of the impurity susceptibility
in Sec.~\ref{sec:susc}, where the coefficient $C_{\rm imp}$ in Eq.~(\ref{fract})
was obtained in an expansion in $\epsilon$ and $r$.

\begin{figure}
\epsfxsize=2.8in
\centerline{\epsffile{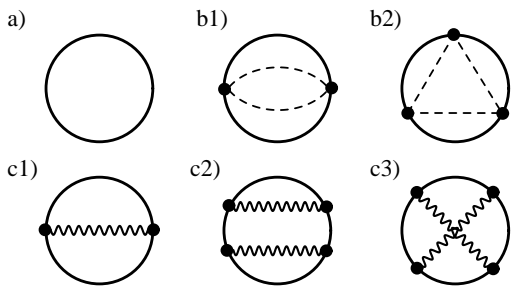}}
\caption{
Feynman diagrams for the impurity partition function of the
physical sector of the Hilbert space, i.e., $\langle\hat{Q}\rangle_\lambda$.
These diagrams directly enter the calculation of the impurity entropy.
Full/dashed/wiggly lines denote $f$/$c$/$\phi$ propagators,
and the dots are the interaction vertices.
a) Unperturbed impurity part.
b) and c) Corrections from the fermionic and bosonic baths.
}
\label{fig:z_dgr}
\end{figure}

The diagrams necessary for the evaluation of the thermodynamic potential are shown
in Fig.~\ref{fig:z_dgr}.
In the following we list the corresponding results;
some details of the evaluation are described in App.~\ref{app:entropy} --
in particular it is important to note that certain {\em disconnected} diagrams
appear in the expansion of the thermodynamic potential.

The coupling to the fermions, Fig.~\ref{fig:z_dgr}b1, yields
\begin{eqnarray}
\Delta S_{\rm imp}^{({\rm f})} = \frac{3j_0^2}{8}
\frac{1}{4 T^2}\int {\rm d}k_1 {\rm d}k_2 \frac{|k_1|^r |k_2|^r}{k_1-k_2}
\frac{k_2}{\cosh^2(k_2/2T)} \,.
\nonumber
\end{eqnarray}
The integrals can be performed in the limit of infinite UV cutoff, with the result
\begin{eqnarray}
\nonumber
\Delta S_{\rm imp}^{({\rm f})} =
\frac{3\pi}{8} \, j_0^2 \, T^{2r} \tan\!\frac{\pi r}{2}\, (2\!-\!2^{1-2r}) \Gamma(2+2r) \zeta(1+2r)
\end{eqnarray}
where $\Gamma(z)$ and $\zeta(z)$ are the Gamma and Riemann Zeta functions.
In the weak-coupling regime the above term goes to zero as $T^{2r}$ for
$T\to0$, thus the entropy at the LM fixed point is $\ln 2$.
However, in the quantum critical region, the dimensionless combination $j_0 T^r$
approaches a universal value at the F-CR fixed point, and
a universal correction to the impurity entropy arises.
This is easily seen by replacing $j_0$ with the renormalized coupling $j$, defined by
$j_0 = j \mu^{-r}$.
Using $(T/\mu)^r = 1 + r \ln (T/\mu) + {\cal O}(r^2)$ and expanding in $r$
one finally obtains:
\begin{equation}
\Delta S_{\rm imp}^{({\rm f})} = \frac{3\pi^2 \ln 2}{8} \, j^2 \, r \,.
\label{simpf}
\end{equation}
Notably, this $j^2$ correction to $S_{\rm imp}$ is suppressed by an additional factor
of $r$.
We have verified that the contributions from the next-order diagram, Fig.~\ref{fig:z_dgr}b2,
proportional to $j^3$, also receive an additional factor of $r$;
thus the term (\ref{simpf}) is the only one to order $r^3$ at the
F-CR fixed point.

The lowest-order bosonic contribution from Fig.~\ref{fig:z_dgr}c1 turns
out to vanish identically for the $g=0$ case, i.e., for non-interacting bosons,
because the contribution of this diagram to the thermodynamic potential
is temperature-independent.
For interacting bosons, we take into account the $T$-dependent mass renormalization
arising from $g$, and obtain using (\ref{mass}) in a similar manner as above:
\begin{equation}
\Delta S_{\rm imp}^{({\rm b_{2,int}})} = - \frac{3\pi^2}{8} \,\gamma^2\, \sqrt{\frac{10\epsilon}{33}} \,.
\end{equation}
For non-interacting bosons, non-vanishing corrections to the entropy
arise only at order $\gamma^4$ from the diagrams shown in
Figs.~\ref{fig:z_dgr}c2, \ref{fig:z_dgr}c3.
With Eq.~(\ref{Omega_disconn}) one finds:
\begin{eqnarray}
\Delta S_{\rm imp}^{({\rm b_4})} &=& -\partial_T
\bigg( (3c2) + (3c3) - (3c1)^2 \bigg) \nonumber\\
&=& - \frac{3\pi^2}{8} \gamma^4 \,.
\end{eqnarray}

With these perturbative results, we can finally evaluate the
residual impurity entropy at the intermediate-coupling fixed points,
to the order available.
\begin{eqnarray}
 \mbox{F-CR}:~S_{\rm imp} &=& \ln 2 \bigg(1 + \frac{3\pi^2}{8}\,r^3\bigg) + {\cal O}(r^5) \,,\nonumber \\
\mbox{FB-CR}:~S_{\rm imp} &=& \ln 2 - \frac{3\pi^2}{16} \sqrt{\frac{10}{33}}\,\,\epsilon^{3/2}
 + {\cal O}(\epsilon,r)^2 \,, \nonumber\\
 \mbox{B-FL}:~S_{\rm imp} &=& \ln 2 - \frac{3\pi^2}{16} \sqrt{\frac{10}{33}}\,\,\epsilon^{3/2}
+ {\cal O}(\epsilon^2) \,.
\label{entr1}
\end{eqnarray}
The last two lines are for interacting bosons \cite{error}.
In principle, the two-loop result for the impurity entropy can be obtained
from the diagrams in Fig.~\ref{fig:z_dgr}
using the methods of Ref.~\onlinecite{ss}; we do not attempt this here.
The entropies for the FB-CR and B-FL fixed points get modified for non-interacting bosons ($g=0$):
\begin{eqnarray}
\mbox{FB-CR}:~S_{\rm imp} &=& \ln 2 - \frac{3\pi^2}{32} \epsilon^2  + {\cal O}(\epsilon^3,r^3) \,, \nonumber\\
 \mbox{B-FL}:~S_{\rm imp} &=& \ln 2 - \frac{3\pi^2}{32} \epsilon^2  + {\cal O}(\epsilon^3) \,.
\label{entr2}
\end{eqnarray}
In App.~\ref{app:largen} we present a large-$N$ calculation \cite{vbs} of the entropy at
the B-FL fixed point, which also yields an answer of the form
$S_{\rm imp} = \ln 2 - {\cal O}(\epsilon^2)$ for small $\epsilon$,
but provides results \cite{entrrem} for all $0\leq \epsilon\leq 2$.

All of the above expressions are consistent with
\begin{eqnarray}
S_{\rm imp}^{\rm F-CR} &\geq& ~~\ln 2~~~\,\, \geq S_{\rm imp}^{\rm B-FL} \,, \nonumber\\
S_{\rm imp}^{\rm F-CR} &\geq& S_{\rm imp}^{\rm FB-CR} \geq S_{\rm imp}^{\rm B-FL} \,;
\label{gtheo}
\end{eqnarray}
these are the conditions that the impurity entropy always decreases in the course
of the renormalization group flow \cite{gtheorem}.
Apparently, in the $\epsilon$ expansion we have
$S_{\rm imp}^{\rm FB-CR} = S_{\rm imp}^{\rm B-FL}$
to the order calculated here, because the coupling
constants $\gamma$ of both fixed points are equal to
one-loop order.
However, it can be expected that in general
$S_{\rm imp}^{\rm FB-CR} > S_{\rm imp}^{\rm B-FL}$ due to the facts
that $\gamma^{\rm FB-CR} < \gamma^{\rm B-FL}$ and $\Delta S_{\rm imp}^{({\rm f})}>0$.

For the F-CR fixed point, the NRG calculations of Ref.~\onlinecite{GBI}
indicated a very small deviation of the impurity entropy from $\ln 2$.
We have repeated these calculations and have found indications for
a correction which scales as $r^3$ at small $r$, however, the accuracy is
not sufficient to reliably determine the prefactor.
Note that the entropy of the fermionic particle-hole symmetric critical fixed point
also approaches \cite{GBI} the value of $\ln 2$
for $r\to \frac{1}{2}$, rendering the corrections to $\ln 2$ tiny for
all $0<r<\frac{1}{2}$.
Interestingly, a similar entropy calculation can be done for
the particle-hole {\em asymmetric} critical fixed point of the pseudogap
Kondo model, where the expansion is performed
around the upper-critical ``dimension'' $r=1$.
The corresponding result\cite{MVLF} is $S_{\rm imp} = \ln 3 - (8/9) (1\!-\!r) \ln 2$,
in good agreement with numerics \cite{GBI}.


\section{$T$ matrix}
\label{sec:tmatrix}

An important quantity in a Kondo model
is the conduction electron $T$ matrix, describing the scattering of
the $c$ electrons off the impurity.
It is useful to define a propagator, $G_T$, of the composite
operator $T_\sigma = f_\sigma^\dagger f_{\sigma'} c_{\sigma'}$,
such that the $T$ matrix is given by $T(\omega) = j_0^2 G_T(\omega)$.
As with the local susceptibility, we
expect a power law behavior of the
$T$ matrix spectral density at the intermediate-coupling fixed points:
\begin{equation}
T(\omega) \propto \frac{1}{\omega^{-r-\eta_T}} \,,
\end{equation}
note that at tree level $G_T \propto\omega^r$ in the present pseudogap
problem.

We now evaluate the lowest-order diagrams of the $T$ propagator  in an
expansion in $\gamma$ and $j$,
introduce a $G_T$ renormalization factor, $Z_T$, and determine
$Z_T$ by minimal subtraction of poles.
The result, given in (\ref{zt}), yields the anomalous exponent
as $\eta_T = -2j + j^2 + 2(\gamma^2-\gamma^4)$.
Evaluating this at the RG fixed points we obtain
\begin{eqnarray}
 \mbox{F-CR}:~\eta_T &=& -2r \,, \\
\mbox{FB-CR}:~\eta_T &=& -2r \,,  \nonumber\\
 \mbox{B-FL}:~\eta_T &=&
\left\{
\begin{array}{ll}
\epsilon - \epsilon^2
   \left( \frac{5}{242}\!+\!\frac{5\pi^2}{66} \right) + {\cal O}(\epsilon^3)
  & (g_0\neq0) \\[8pt]
\epsilon
  & (g_0=0)
\end{array}
\right.
.   \nonumber
\end{eqnarray}
Remarkably, the two-loop corrections for F-CR and FB-CR have cancelled.
As announced in Ref.~\onlinecite{MVMK}, this cancellation holds to all
orders.
The argument parallels the one for anomalous exponent of the local
susceptibility of {\em non-interacting} bosons, given in
Refs.~\onlinecite{bfknew,science}.
For fermions {\em without} a relevant self-interaction, all diagrams
for $G_T$ can be
obtained from the first diagram with {\em full} $j$ vertices and
{\em full} $f$ electron propagators.
This gives the relation $Z_T^{-1} = Z_j^2 / Z_f^2$;
note that the corresponding relation for the local susceptibility of
non-interacting bosons is $Z_\chi^{-1} = \widetilde{Z}_\gamma^2 / Z_f^2$.
Using the definition of $Z_j$ (\ref{jdef}) and taking the derivative w.r.t. $(\ln\mu)$
at fixed bare $j_0$, we obtain $0=-r + \beta(j)/j - \eta_T/2$.
Thus, at any RG fixed point with finite $j^\ast$ we
find the exact result
\begin{equation}
\eta_T=-2r ~~\Rightarrow~~
T(\omega) \propto \omega^{-r}
\label{exact}
\end{equation}
which applies to the F-CR and FB-CR fixed points.
For F-CR it is in perfect agreement with NRG results \cite{bulla,MVRB}.

At finite temperatures, hyperscaling again implies a
scaling form of the $T$ matrix, which can be written as:
\begin{equation}
T(\omega,T)
= \frac {{\cal B}_T} {\omega^r} \, \Phi_T \!\left(\frac{\omega}{T}\right)
\,,
\label{imt}
\end{equation}
valid at both the F-CR and FB-CR fixed points.

Away from criticality, scaling is also obeyed as function of the distance
from the critical point, similar to Eq.~(\ref{chiscal2}).
We note that in the LM regime, bare perturbation theory is applicable
and gives $T(\omega) \propto \omega^r$.
Then, for $j_0\lesssim j_{0{\rm c}}$
in the pseudogap Kondo problem, one finds a crossover from
$T(\omega)\propto\omega^{-r}$ for $\omega\gg T^\ast$ to
$T(\omega)\propto\omega^{r}$ for $\omega\ll T^\ast$,
in agreement with NRG results \cite{bulla}.

\begin{figure}
\epsfxsize=3.4in
\centerline{\epsffile{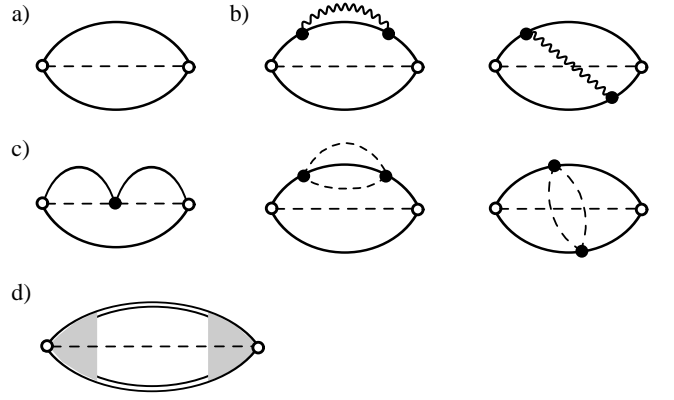}}
\caption{
Feynman diagrams for the Green's function $G_T$ entering the conduction
electron $T$ matrix.
Notation is as in Fig.~\protect\ref{fig:z_dgr};
the double line is the full $f$ propagator, the shaded area the full $j$ vertex,
and the open dots are the external sources.
a) Bare $T$ matrix.
b) and c) Corrections from the fermionic and bosonic baths.
d) Full $T$ matrix.
}
\label{fig:t_dgr}
\end{figure}

For transport experiments in a pseudogap system with dilute magnetic impurities,
we can estimate the impurity-induced resistivity $\rho_{\rm imp}$
by evaluating the conduction electron bubble with the lowest-order self energy.
The above implies a divergence
$\rho_{\rm imp} \propto T^{-r(1-r)}$ at criticality
(with logarithmic corrections at both $r=0$ and $r=1$),
whereas $\rho_{\rm imp}$ vanishes according to
$\rho_{\rm imp} \propto T^{r(1-r)}$ away from criticality.
Here, the $(1-r)$ factor in the exponent arises from the momentum
integration
(consistent with a resistivity being independent of the scattering
rate at $r=1$, Ref.~\onlinecite{palee}).
Thus, for $0<r<1$ and sufficiently close to the critical point, one expects a
characteristic peak in the temperature dependence of $\rho_{\rm imp}$,
located at the crossover temperature $T^\ast$.


\section{Dynamical mean-field theory}
\label{sec:appl}

Magnetic ordering transitions in certain heavy-fermion compounds,
like \cecuau and \ybrhsi,
have been proposed to be accompanied by the breakdown of Kondo
screening at criticality \cite{edmft,bgg}.
One particular scenario \cite{edmft} employs an extended DMFT description
for this transition, where the original Kondo lattice model
is mapped onto a self-consistent version of the Fermi-Bose
Kondo model, and the critical point of the bulk system is
mapped onto the BF-CR critical point of the impurity
model.
Recent numerical calculations have found the existence
of such a transition in the extended DMFT for an anisotropic
Kondo lattice model \cite{sinum}
(however, competing claims have also been published \cite{gabi_edmft}).

In the following,
we want to focus on a property of such a locally-critical
point which has not received much attention so far, namely
the entropy and specific heat near the critical point.
In principle, our statements will apply to all quantum critical
points in DMFT where the effective impurity model is itself critical
at the bulk transition.
(Note that this does {\em not} seem to be the case for the
much-studied Mott metal--insulator transition in the one-band Hubbard
model\cite{dmft}.)

To be specific, let us consider the thermodynamic potential of an Anderson
lattice model, treated within extended DMFT
(equations for a $t-J$ model are similar).
It has been shown in Ref.~\onlinecite{haule} that the lattice potential (per site)
can be written as the sum of impurity and bath contributions,
\begin{eqnarray}
\Omega &=& \Omega_{\rm imp} + \Omega_c + \Omega_{\phi} \,, \label{entrlat} \\
\Omega_c &=& T\sum_{\omega_n}
  \frac{1}{N_s} \sum_k \ln [G_c(k,i\omega_n)/G_{\rm loc}(i\omega_n)] e^{i\omega_n 0^+} \,, \nonumber\\
\Omega_\phi &=& \frac{T}{2} \sum_{\nu_n}
  \frac{1}{N_s} \sum_k \ln [D_\phi(k,i\nu_n)/\chi_{\rm loc}(i\nu_n)] e^{i\nu_n 0^+} \nonumber \,,
\end{eqnarray}
where $G_c(k,i\omega_n)$ and $D_\phi(k,i\nu_n)$ are the fermionic and
bosonic bath Green's functions, respectively, and
$N_s$ is the number of lattice sites.
$G_{\rm loc}$ and $\chi_{\rm loc}$ are the (local) impurity
electron propagator and susceptibility -- in our notation,
$G_{\rm loc}$ is the propagator of the $f$ fermions.
The lattice entropy (per site) is simply $-\partial_T\Omega$.
The temperature derivatives of the two terms $\Omega_c$ and $\Omega_{\phi}$
vanish in the zero-temperature limit:
note that they just represent contributions from systems of free fermions
or bosons, where
the respective densities of states are (at most) logarithmically singular
within the locally-critical point scenario in two space dimensions \cite{edmft}.

Thus, the entropy of the impurity model, $-\partial_T\Omega_{\rm imp}$,
is the only contribution to the total entropy of the lattice model
which survives as $T\to 0$.
The results of Sec.~\ref{sec:entr} suggest that the critical point
of the Fermi-Bose Kondo model, BF-CR, generically has a finite residual
entropy \cite{entrrem},
with the perturbative result given in Eqs.~(\ref{entr1},\ref{entr2}).
According to (\ref{entrlat}), this implies an extensive entropy of the
Kondo lattice model described by extended DMFT, which should be
present at the $T=0$ critical point as well as in the low-temperature
limit of the quantum critical region.
As the ground state entropies of the stable phases
(heavy Fermi liquid and antiferromagnet for the case of the Kondo lattice)
are non-extensive, a huge anomaly in the
specific heat, $C = T (\partial S/\partial T)$, should be observed
upon crossing the phase diagram at low temperatures above
the quantum critical point.
To our knowledge, this has not been observed experimentally.
Furthermore, it can be considered unlikely that a quantum critical point
of a lattice model displays an extensive residual entropy, as
such a point would be unstable to any perturbation.
(Note that a generic lattice quantum critical point has only {\em one}
unstable direction in the RG flow, and vanishing residual entropy.)

Finally, as detailed in Ref.~\onlinecite{MVMK}, the DMFT
self-consistency equation for the fermionic bath at local criticality can
only be fulfilled with a fermionic spectrum with $r=0$.
Therefore, the local critical point displays at most logarithmic corrections
to a metallic local density of states, and one can thus expect
a logarithmic $T$-dependence of the resistivity.
A discussion of the resistivity away from criticality requires
to consider the fully self-consistent problem, and is beyond the scope
of this paper.


\section{Discussion and conclusions}

We have calculated critical properties of a Fermi-Bose Kondo model,
describing an impurity spin coupled to both fermionic quasiparticles with a
pseudogap DOS and collective mode bosons,
using renormalization group techniques.
In particular, we have determined the impurity susceptibilities and
residual entropies at different intermediate-coupling fixed points.

Our results for the phase transition in the pseudogap Kondo model
can be compared with numerical results obtained by high-accuracy NRG
calculations.
The excellent agreement strongly suggests that the
renormalized perturbation expansion approach for critical properties is
reliable in the context of impurity quantum phase transitions
(provided that the fixed point is in the applicability range
of the renormalization group procedure),
and gives numerically accurate results for a {\em finite}
range of the (small) parameters $r$, $\epsilon$.

Remarkably, the physics found in the present models is
somewhat different from the one of the two-channel Kondo effect.
In the two-channel problem, a number of thermodynamic
properties vanish in the strict scaling limit, and
corrections to scaling yield the leading low-temperature observables.
In contrast, for the intermediate-coupling fixed points considered
here, the predictions of naive scaling apply.
The reason for this difference apparently is the finite tree-level
scaling dimension of the impurity couplings in the present models,
whereas the coupling in the metallic two-channel problem is
marginal.

As can be seen from the perturbative results presented in this paper,
the prefactors of most of the higher-order corrections in the $\epsilon$ expansion of
physical quantities are not small, and thus no quantitatively accurate
numbers can be obtained for $\epsilon=1$.
It may be possible to use certain resummation techniques of the
perturbative expansion -- this is beyond the scope of this work.
Nevertheless, we expect that the qualitative physics is captured
by the present study, which is supported by the large-$N$ calculations
in Ref.~\onlinecite{vbs} and App.~\ref{app:largen} for the purely bosonic
Kondo problem.

On the single-impurity application side,
the models discussed here can describe
magnetic impurities in strongly correlated materials, where the interplay of
quasiparticles and collective low-energy excitations is a central
ingredient.
As has been argued elsewhere \cite{tolya,MVRB,MVMK}, the
pseudogap Kondo model appears relevant for impurity moments
in high-temperature superconductors.
Detailed calculations of the local density of states in the
vicinity of the impurity \cite{tolya,MVRB} have been compared with
scanning tunneling microscopy results \cite{seamuszn} in Zn-doped \bscco.
In addition, the interplay with bulk spin fluctuations possibly
explains the large doping dependence of the Kondo temperature
across the phase diagram of \ybco \cite{bobroff1,bobroff2},
as has been explicitely demonstrated in calculations using the pseudogap
Fermi-Bose Kondo model in Ref.~\onlinecite{MVMK}.
In this case, the inclusion of the bosonic self-interaction is
important, as fermionic quasiparticles and spin fluctuations are decoupled
in the low-energy physics of a $d$-wave superconductor, and Landau damping
is absent \cite{vbs}.

Finally we have discussed the issue of the ground-state entropy
in scenarios of local quantum criticality, described by
dynamical mean-field theory.
We found that the finite residual entropy of the critical impurity
model, found in Sec.~\ref{sec:entr} \cite{entrrem}, leads to an extensive
ground-state entropy of the lattice model,
which appears unlikely to be realized in a finite-dimensional
system.
One can conclude that the dynamics at a heavy-fermion critical point,
which is associated with the breakdown of Kondo screening,
{\em cannot} be entirely local, as described by dynamical mean-field
approaches.
In other words, at lowest temperatures the entropy has to be quenched
by non-local correlations.
Versions of DMFT can still provide a reasonable description of the
physics at intermediate energies and temperatures.


\acknowledgments

We thank R. Bulla, L. Fritz, A. Polkovnikov, and S. Sachdev
for discussions and collaborations on related work,
as well as P. Coleman, S. Florens, O. Parcollet, A. Rosch, and Q. Si for helpful
discussions.
MV is particularly grateful to C. Lorenz for exposing
the role of Eq. (B1).
This research was supported by the DFG through the Center
for Functional Nano\-struc\-tures Karls\-ruhe.


\appendix

\section{Field-theoretic RG to two loops}
\label{app:twoloop}

The renormalization group to two-loop order is conveniently
done in the field-theoretic scheme, employing dimensional regularization and
minimal subtraction of poles.
Using the conventions of Ref.~\onlinecite{bgz}, we introduce renormalized
fields and dimensionless couplings for the bulk bosons
\begin{eqnarray}
\phi_\alpha &=& \sqrt{Z} \, \phi_{R\alpha} \,,  \\
g_0 &=& \frac{\mu^\epsilon Z_4} {Z^2 S_{d+1}} g \label{gdef}
\end{eqnarray}
and for the impurity
\begin{eqnarray}
f_\sigma &=& \sqrt{Z_f} \, f_{R\sigma} \,, \\
\gamma_0 &=& \frac{\mu^{\epsilon/2} \widetilde{Z}_{\gamma}}{Z_f\sqrt{Z\widetilde{S}_{d+1}}} \gamma
\,, \label{gammadef} \\
j_0 &=& \frac{\mu^{-r} Z_j}{Z_f} j \,; \label{jdef}
\end{eqnarray}
here $\mu$ is a renormalization
energy scale, and
\begin{eqnarray}
S_d &=& \frac{2}{\Gamma(d/2) (4 \pi)^{d/2}} \,, \nonumber\\
\widetilde{S}_d &=& \frac{\Gamma(d/2-1)}{4 \pi^{d/2}}.
\label{sd}
\end{eqnarray}
No renormalizations are needed for the fermions as their self-interaction
is assumed to be irrelevant in the RG sense.

The needed diagrams have been evaluated in
Refs.~\onlinecite{bfknew} and \onlinecite{issp} and will not be repeated here.
Minimal subtraction of poles yields the renormalization factors for
the bulk bosons \cite{bgz}
\begin{eqnarray}
Z &=& 1 - \frac{5 g^2}{144 \epsilon} \,, \nonumber \\*
Z_4 &=& 1 + \frac{11g}{6 \epsilon} + \left(
\frac{121}{36\epsilon^2} - \frac{37}{36\epsilon} \right) g^2
\label{eps10}
\end{eqnarray}
and for the impurity
\begin{eqnarray}
Z_f &=& 1 +
  \frac{3\,j^2}{16\,r} -
  \frac{3\,{\gamma }^2}{4\,\epsilon } +
  \left(-\frac{15}{32\,{\epsilon }^2} +
  \frac{3}{8\,\epsilon } \right)\gamma^4  \,, \nonumber\\
\widetilde{Z}_{\gamma} &=& 1 -
  \frac{j^2}{16r} +
  \frac{{\gamma }^2}{4\epsilon } +
  \left( \frac{9}{32{\epsilon }^2} - \frac{1}{8\epsilon } \right)\gamma^4  +
  \frac{5g{\pi }^2{\gamma }^2}{72\epsilon } \,, \nonumber \\
Z_j &=& 1 +
  \frac{j}{r} +
  \left(\frac{1}{r^2} - \frac{1}{16r}\right) j^2+
  \frac{{\gamma }^2}{4\,\epsilon } +
  \left(\frac{9}{32{\epsilon }^2} - \frac{1}{8\epsilon } \right) \gamma^4
  \nonumber\\ &&~~~~~~+
     \left( \frac{1}{4\,\epsilon } +
       \frac{1}{\epsilon -r} \right)   \frac{j\,{\gamma }^2 }{r}
\,.
\end{eqnarray}
The RG beta functions can now be evaluated by taking the $\mu$
derivatives of (\ref{gdef}), (\ref{gammadef}), and (\ref{jdef})
at fixed values of bare coupling constants.
For the bulk coupling we find the known result:
\begin{eqnarray}
\beta(g) \equiv \mu \frac{{\rm d}g}{{\rm d}\mu} \bigg|_{g_0}
&=& - \epsilon g + \frac{11 g^2}{6} - \frac{23 g^3}{12}
\label{eps18}
\end{eqnarray}
which has the stable fixed point
\begin{equation}
g^{\ast} = \frac{6 \epsilon}{11} + \frac{414 \epsilon^2}{1331}\,.
\end{equation}
The impurity couplings are governed by the following beta functions:
\begin{eqnarray}
\beta(\gamma) &=& -\frac{\epsilon\gamma}{2} + \gamma^3 - \gamma^5 + \frac{5g^2\gamma}{144}
+ \frac{5\pi^2 g\gamma^3}{36}
+ \frac{j^2\gamma}{2} \,, \nonumber \\
\beta(j) &=& rj - j^2 + \frac{j^3}{2} + j\gamma^2 - j\gamma^4 \,.
\label{twoloopbeta}
\end{eqnarray}
Using these beta functions one obtains the fixed point values of the
couplings quoted in Sec.~\ref{sec:FP}.

The anomalous field dimension $\eta_f$ is obtained from
\begin{equation}
\eta_f = \mu \frac{{\rm d} \ln Z_f}{{\rm d}\mu} \bigg|_{g^\ast,\gamma^\ast,j^\ast} \,,
\end{equation}
where the derivative is evaluated at the RG fixed point.

To calculate the anomalous exponent of the local susceptibility,
a renormalization factor, $Z_\chi$, for the two-point correlations
of the impurity spin has to be introduced.
The required diagrams have been evaluated in
Refs.~\onlinecite{bfknew} and \onlinecite{vbs};
the result for $Z_\chi$ within the minimal subtraction scheme is
\begin{equation}
Z_\chi = 1- \frac{2\gamma^2}{\epsilon} + \frac{\gamma^4}{\epsilon} + \frac{j^2}{2r} \,.
\label{zchi}
\end{equation}
Employing
\begin{equation}
\eta_\chi = \mu \frac{{\rm d} \ln Z_\chi}{{\rm d}\mu} \bigg|_{g^\ast,\gamma^\ast,j^\ast}
\end{equation}
one obtains the $\eta_\chi$ values quoted in Sec.~\ref{sec:susc}.

Similarly, the renormalization factor for the $T$ matrix, discussed in Sec.~\ref{sec:tmatrix},
is obtained as
\begin{eqnarray}
Z_T &=&
1 - \frac{2 j}{r} + \frac{j^2}{2r} +\frac{j^2}{r^2} - \frac{2\gamma^2}{\epsilon}
+ \frac{\gamma^4}{\epsilon}  \nonumber\\ &&~~~~~~+
     \left( \frac{4}{\epsilon } -
       \frac{2}{r} \right)   \frac{j\,{\gamma }^2 }{r-\epsilon} \,.
\label{zt}
\end{eqnarray}


\section{Role of $\langle\hat{Q}\rangle$}
\label{app:denom}

Here we briefly discuss the role of the denominator appearing in
the equation for observables, (\ref{obs}), within the pseudo-fermion
technique employed here.
(Note that a similar denominator also appears in the formalism used
in Ref.~\onlinecite{vbs}.)

For a renormalization group procedure which directly calculates renormalizations
of vertices and propagators (as the momentum shell method), the
denominator does formally not appear.
In contrast, when calculating true observables, as e.g. done in Ref.~\onlinecite{vbs},
$\langle\hat{Q}\rangle$ needs to be taken into account.
Importantly, approximating the denominator with its leading contribution,
$2 \exp(-\beta\lambda)$, is in general not sufficient, as the denominator
contains -- as the numerator of (\ref{obs}) -- contributions from all
orders in an expansion in the non-linear couplings.

However, in the calculation of critical exponents using, e.g., the minimal
subtraction scheme, one only needs to keep track of contributions which
develop poles in $\epsilon$ after dimensional regularization, i.e., are
logarithmic at the marginal dimension.
Typically, the denominator does {\em not} develop such poles,
which can be seen by comparing power counting for response functions,
i.e., second derivatives of the thermodynamic potential, and for the potential
itself.
Therefore, for the calculation of anomalous exponents it is permissible to
ignore the perturbation expansion of $\langle\hat{Q}\rangle$, as done
in Ref.~\onlinecite{bfknew}.

In contrast, when calculating a quantity which does {\em not} develop an anomalous
power law, like $\chi_{\rm imp}$ in the present impurity problems,
all poles cancel, and one has to collect contributions non-singular in $\epsilon$.
Those also arise from a perturbative expansion of the denominator,
and typically cancel corresponding terms from the numerator.
Explicitly, we have to lowest order:
\begin{eqnarray}
\langle\hat{Q}\rangle = 2 \exp(-\lambda\beta) \bigg(
1 - \frac{3}{8T} \gamma_0^2 \int \frac{{\rm d}^d k}{(2\pi)^d} \,\frac{1}{\epsilon_k^2}
\bigg) \,;
\end{eqnarray}
to this order only the bosonic part enters, and
$\epsilon_k$ is the boson dispersion appearing in
$D_\phi(k,\omega_n) = (\omega_n^2+ \epsilon_k^2)^{-1}$.

Finally, we note that in certain impurity problems the real part of the
propagator renormalization requires the introduction of a suitable
counter-term -- physically this reflects the shift of the
phase transition point, arising from the real part of
a self-energy.
In the present case no counter-terms are necessary.


\section{Perturbative expansion for the impurity entropy}
\label{app:entropy}

In this appendix we describe the perturbation theory for the
impurity part of the thermodynamic potential.
The impurity spin is represented by $f$ pseudo-fermions as
above, together with the chemical potential in the limit
$\lambda\to\infty$.
We shall illustrate that the correct treatment of this
limit has non-trivial consequences for the higher-order
diagrammatic expansion of the entropy.

The starting point is the partition function in the physical
sector of the Hilbert space ($\hat{Q}=f_\nu^\dagger f_\nu=1$), which can be
written as
\begin{equation}
\frac{Z_{\rm imp}}{Z_{{\rm imp},0}} =
\lim_{\lambda\to\infty}
\frac{\langle\hat{Q}\rangle_\lambda}{\langle\hat{Q}\rangle_{\lambda,0}}
\label{Zimp}
\end{equation}
Here, $\langle \ldots \rangle_{\lambda}$ is an expectation
value with the full Hamiltonian in the presence of $\lambda$
as above, and $\langle \ldots \rangle_{\lambda,0}$ is
the expectation value in the absence of the coupling between
impurity and baths.
The above expression is easily understood:
for $\lambda\to\infty$, ${\langle\hat{Q}\rangle_\lambda}$ is the
partition in the physical sector times $\exp(-\lambda\beta)$, i.e.,
the ``factor'' $\hat{Q}$ suppresses the $\hat{Q}=0$ part of the Hilbert space.
Further, in this limit
$\langle\hat{Q}\rangle_{\lambda,0} = 2 \exp(-\lambda\beta)$,
and the unperturbed impurity part of the partition function is
just $Z_{{\rm imp},0} = 2$ for a spin-$\frac{1}{2}$ impurity.
Importantly, the limit $\lambda\to\infty$ in $Z_{\rm imp}$
suppresses the contributions from disconnected diagrams,
consisting of more than one loop of $f$ pseudo-fermions,
because those are smaller by further powers of $\exp(-\lambda\beta)$.

The thermodynamic potential is given by
\begin{equation}
\Omega_{\rm imp} - \Omega_{{\rm imp},0} =
- T \ln \frac{Z_{\rm imp}}{Z_{{\rm imp},0}}
\,.
\label{omegaimp}
\end{equation}
The partition function (\ref{Zimp}) can now be expanded in the
impurity couplings using standard diagrammatics, and then the
log in (\ref{omegaimp}) is expanded to the required order.
There is now {\em no} cancellation of disconnected
diagrams in $\Omega_{\rm imp}$, because those diagrams do not
appear in $Z_{\rm imp}$ -- this reflects the fact that we are
not expanding around a system of free particles here.

For the bosonic impurity problem, we have
\begin{equation}
\frac{Z_{\rm imp}}{Z_{{\rm imp},0}} =
1 + \frac{\gamma_0^2}{2} D^{(2)} + \frac{\gamma_0^4}{4} D^{(4)} + \ldots
\end{equation}
where $D^{(n)}$ is the sum of all $n$-th order different {\em connected}
diagrams; $D^{(2)}$ and $D^{(4)}$ are shown in Fig.~\ref{fig:z_dgr}c.
Thus,
\begin{equation}
\frac{\Omega_{\rm imp} - \Omega_{{\rm imp},0}}{- T} =
\frac{\gamma_0^2}{2} D^{(2)} + \frac{\gamma_0^4}{4} D^{(4)} - \frac{\gamma_0^4}{8} {D^{(2)}}^2 +
\ldots \,,
\label{Omega_disconn}
\end{equation}
i.e., the expansion for the thermodynamic potential contains
disconnected diagrams!

The entropy is obtained from $\Omega_{\rm imp}$ by
$S_{\rm imp} = - \partial_T \Omega_{\rm imp}$.
Power counting shows that $S_{\rm imp} = \ln 2$ as $T\to 0$
in the cases where the fixed point value of the impurity coupling is zero
(as is the case, e.g., in the Bose-Kondo model for $d>3$ space dimensions).
Then, only subleading corrections arise from the coupling to the bath.
In contrast, in the case of a finite renormalized coupling $j^\ast$ or $\gamma^\ast$,
bare perturbation theory is infrared divergent, as for standard
problems below their upper-critical dimension.
Proceeding in the scheme of renormalized perturbation theory,
the perturbative result (to lowest non-trivial order)
can be expressed in terms of the renormalized couplings.
At the RG fixed point, the couplings can then be replaced by their
fixed-point values, and the result for the entropy is finite
as $T\to 0$.

The impurity part of the thermodynamic potential
diverges with the cutoff, i.e., we have
$\Omega_{\rm imp} = E_{\rm imp} - T S_{\rm imp}$,
where $E_{\rm imp}$ is the non-universal (cutoff-dependent) impurity contribution
to the ground-state energy.
However, the impurity entropy $S_{\rm imp}$ is fully universal, and
the UV cutoff can be sent to infinity {\em after} taking the
temperature derivative of $\Omega_{\rm imp}$.


\newcommand{\bb}  {{f}}         
\newcommand{\ttp} {{\bar t}}    
\newcommand{\ssp} {{\bar s}}    
\newcommand{\ttd} {{t}}         
\newcommand{\KK}  {{K}}         
\newcommand{\JJ}  {{J}}         
\newcommand{\KKK} {{\tilde K}}  
\newcommand{\OM}  {{\epsilon}}  
\newcommand{\MM}  {m}           
\newcommand{\DD}  {\Delta}      
\newcommand{\LL}  {{\lambda_0}} 
\newcommand{\coup} {s}          

\section{Large-$N$ theory of the Bose Kondo problem -- Entropy}
\label{app:largen}

As the perturbative method employed in the body of the paper
is restricted to small $\epsilon = 3-d$, alternative approaches
are desirable which can provide results for all $\epsilon$.
In Ref.~\onlinecite{vbs} a dynamic large-$N$ approach to the
bosonic Kondo problem has been developed.
Here we shall use this approach to evaluate the impurity entropy
at the B-FL fixed point in arbitrary space dimension $d$,
which complements the results of Sec.~\ref{sec:entr}.

We begin by summarizing the large-$N$ theory of Ref.~\onlinecite{vbs}.
The impurity spin symmetry is generalized from SU(2) to SU($N$).
The spin operator is represented by auxiliary fermions $\bb_\nu$
($\nu=1, ..., N$),
and a chemical potential $\LL$ is introduced to enforce the constraint
$\bb_\nu^\dagger \bb_\nu = N q_0$.
We will restrict the considerations to $q_0=1/2$; this choice has the advantage
of preserving particle-hole symmetry, such that the chemical potential $\LL$
is actually zero for all temperatures.
The Hamiltonian is given by the natural large-$N$ generalization of
${\cal H}_{\rm imp} + {\cal H}_\phi$ of Sec.~\ref{sec:models}, where
the $\phi$ field now has $N^2-1$ components;
for details see Ref.~\onlinecite{vbs}.
Taking the limit $N\to\infty$ results in a dynamic saddle point \cite{OSPG},
and the impurity physics is captured by a self-consistent integral
equation,
\begin{equation}
\Sigma_{\bb}(\tau)=
- \gamma_0^2 D_\phi(\tau) G_{\bb}(\tau) \,,
\label{ncasigma}
\end{equation}
where $D_\phi$ is the local spin fluctuation propagator (\ref{eps2}), and
the self-energy $\Sigma_{\bb}$ is defined by:
\begin{equation}
\label{defsigma}
G_{\bb}^{-1}(i\omega_n) = i\omega_n - \LL - \Sigma_{\bb}(i\omega_n) \,.
\end{equation}
Here $G_{\bb}(\tau) = -\langle {\rm T} \bb(\tau) \bb^\dagger(0) \rangle$
is the {\em full} auxiliary fermion Green's function in standard notation.
Clearly, (\ref{ncasigma}) represents a summation of all self-energy diagrams
with non-crossing boson lines, somewhat similar to the non-crossing approximation
known from fermionic Kondo physics \cite{hewson,OSPG}.
In writing down the above equations we have assumed full SU($N$) symmetry and dropped
the corresponding SU($N$) indices.
Finally, $\LL$ is determined by the constraint equation:
\begin{eqnarray}
\label{eqlambda}
G_{\bb}(\tau=0^-) =
{1 \over\beta  }
\sum_{n} G_{\bb} (i\omega_n) e^{i\omega_n 0^+}
=\,N q_0
,\, q_0 = \frac{1}{2}
\nonumber
\end{eqnarray}
which is designed to satisfy the contraint $\bb_\nu^\dagger \bb_\nu = N q_0$
on average.

In the following we concentrate on the bulk critical point, $s=s_c$, where
the bosonic spectrum is gapless,
$\rho_\phi(\omega) \propto {\rm sgn}(\omega) |\omega|^{1-\epsilon}$.
It has been established in Ref.~\onlinecite{vbs} that
the $T=0$ result for the fermionic Green's function
obeys:
\begin{eqnarray}
\label{longtime}
G_\bb(\tau) \sim
\frac{1}{\tau^{\epsilon/2}}
\,,\quad
\rho_\bb(\omega) \sim
|\omega|^{\epsilon/2 -1}
\:.
\label{gcp}
\end{eqnarray}
in the long-time or low-energy limit;
here $\rho_\bb(\omega)$ is the spectral density of $G_\bb$.
For the case of non-interacting bosons it is possible to
obtain an analytical result for the finite-temperature scaling
form of $G_f$,
\begin{eqnarray}
G_\bb(\tau) &=& - A \,\frac{T^{\epsilon/2}}{\gamma_0^\epsilon}
\left({{\pi}\over{\sin\pi (\tau T)}}\right)^{\epsilon/2}
\label{gfscal} \,,
\end{eqnarray}
which is dictated by conformal invariance \cite{OSPG};
$A$ is an amplitude prefactor depending on $\epsilon$.

The local susceptibility in the large-$N$ limit,
\begin{equation}
\chi_{\text{loc}} =
- \int_0^\beta {\rm d}\tau \: G_\bb(\tau) G_\bb(-\tau)
\:,
\end{equation}
can be evaluated directly at the critical point, $s=s_c$,
with the result $\chi_{\rm loc} \propto T^{-1+\epsilon}$.
Thus we have the large-$N$ result
\begin{equation}
\eta_\chi = \epsilon \,,
\end{equation}
which coincides with the exact result (\ref{exact_etachi})
for non-interacting bosons in the SU(2)-symmetric Bose Kondo problem.

\begin{figure}[t]
\epsfxsize=3.2in
\centerline{\epsffile{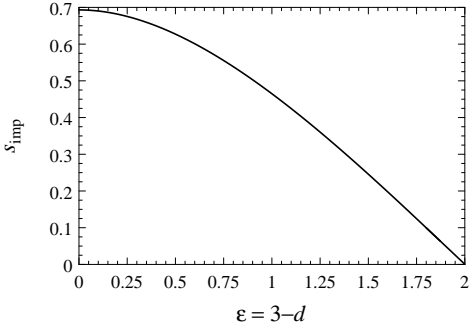}}
\caption{
Large-$N$ result (\protect\ref{simp_pg}) for the
residual impurity entropy (per flavor) at the B-FL fixed point,
valid for a bath of non-interacting bosons in $(3-\epsilon)$
space dimensions.
}
\label{fig:largen}
\end{figure}

Now we proceed with the calculation of the residual impurity entropy,
following the method by Parcollet {\em et al.} \cite{OSPG} for the
fermionic NCA.
At the saddle point, the impurity part of the free energy per flavor
can be written as:
\begin{equation}
f_{\rm imp} = q_0 \LL + T\sum_n \ln G_\bb(i\omega_n)
- \int_0^\beta \!\!\!\! {\rm d}\tau \: \Sigma_\bb(\tau) G_\bb(-\tau) \,.
\label{fimp}
\end{equation}
For particle-hole symmetry, $q_0=1/2$, we have $\LL=0$.
Furthermore, the last term in (\ref{fimp}) behaves
as ${\rm const} + {\cal O}(T^{1+\epsilon/2})$, and thus
does not contribute to the zero-temperature entropy.
It remains to evaluate the contribution from ${\rm Tr}\,\ln\,G_f$.
As shown in Ref.~\onlinecite{OSPG}, it can be written as:
\begin{eqnarray}
f_{\rm imp} =
\frac{1}{\pi} \int_{-\infty}^\infty \!\!{\rm d}\omega \,
n_f(\omega) \left[
\arctan \!\bigg(\frac{G_f'(\omega)}{G_f''(\omega)}\bigg) -\frac{\pi}{2}
\right] ,
\end{eqnarray}
with $G_f'$, $G_f''$ being the real and imaginary part of the
finite-temperature retarded fermion Green's function, and $n_f$ is the
Fermi factor.

In the following we restrict our attention to the case of non-interacting
bosons which allows to make analytical progress.
The temperature derivative of the above expression has to be taken
using the scaling form (\ref{gfscal}) of $G_f$.
After straightforward algebra in analogy to Sec.~VI of Ref.~\onlinecite{OSPG}
we find for the impurity entropy (per flavor):
\begin{eqnarray}
s_{\rm imp} =
- \frac{2}{\pi} \int_0^1 \!\!\frac{{\rm d}u}{1-u^2}
\left[
u \arctan(u t) -\arctan\,t
\right] \,,
\end{eqnarray}
where the substitutions $1/t = \tan (\pi\epsilon/4)$ and $u = \tanh(\omega/2T)$
have been employed.
The above result can be re-written as
\begin{eqnarray}
s_{\rm imp} = \ln 2 -
\frac{2}{\pi} \int_0^{1/t} \!\!{\rm d}v \,
\frac{v \arctan(1/v)}{1+v^2} \,.
\label{simp_pg}
\end{eqnarray}
This represents our large-$N$ result for the impurity entropy at the B-FL
fixed point for non-interacting bosons.
Numerical evaluation of the integral yields $s_{\rm imp}$ as shown in
Fig.~\ref{fig:largen}.
For small $\epsilon$, where $1/t = \pi\epsilon/4$, the correction
to the $\log 2$ entropy is quadratic in $\epsilon$,
which is similar to the SU(2) result in (\ref{entr2}).
In particular, we see that $s_{\rm imp}$ is {\em finite}
for all $0\leq\epsilon<2$, with the numerical value
\begin{equation}
s_{\rm imp}(d=2) = 0.4648477
\end{equation}
at $\epsilon=1$.
(Note that, e.g., the local susceptibility is only singular
for $0\leq\epsilon\leq1$.)
$s_{\rm imp}$ vanishes as $d\to 1$, i.e., at the lower critical
dimension of the bulk theory, where $G_f$ is no longer singular --
this is consistent with the considerations of
Ref.~\onlinecite{impnlsm}.

Unfortunately, the present large-$N$ theory cannot be easily generalized to
the Fermi-Bose Kondo model with a single fermionic screening channel.
Instead, the natural dynamic large-$N$ formulation of the fermionic
sector leads to a multichannel Kondo model \cite{OSPG}, which, however,
does not display a strong-coupling fixed point with vanishing residual
entropy.



\begin{thebibliography}{}
\bibitem{error}
The last line in Eq.~(\ref{entr1}) corrects an error in the entropy
expression given in Ref.~\onlinecite{science}, where a temperature
derivative was missed.
\bibitem{entrrem}
Based on the $\epsilon$ expansion we cannot exclude that
the impurity entropy at the FB-CR fixed point is zero for
$\epsilon=1$, while being finite for small $\epsilon$.
However, this seems unlikely to us because we see no indications for
changes in the fixed-point structure \protect\cite{impnlsm} in the range $0<\epsilon<2$,
at least for interacting
bosons $g \neq 0$; the case $g = 0$ requires further investigation.
Furthermore, the large-$N$ calculation in App.~\ref{app:largen}
shows that the entropy for the bosonic Kondo problem is
finite for all $0<\epsilon<2$,
while -- interestingly -- the local susceptibility
is only singular for $0<\epsilon<1$ due to $\eta_\chi=\epsilon$
(as at the BF-CR fixed point).
\end{thebibliography}
\end{document}